\documentclass{ws-ijmpe}

\usepackage{epsfig}

\usepackage{color}

\begin{document}

%\markboth{B.~M\"uller, A.~Sch\"afer}{Entropy Creation in Relativistic Heavy Ion Collisions}

\catchline{}{}{}{}{}
%%%%%%%%%%%%%%%%%%%%%%%%%%%%%%%%%%%%%%%%%%%%%%%%%%%%

\title{ENTROPY CREATION\\ IN RELATIVISTIC HEAVY ION COLLISIONS}

\author{\footnotesize BERNDT M\"ULLER}

\address{Department of Physics and CTMS, Duke University, Durham, NC 27708, USA\\ 
muller@phy.duke.edu}

\author{ANDREAS SCH\"AFER}

\address{Institut f\"ur Theoretische Physik, Universit\"at Regensburg, D-93040 Regensburg, Germany\\
andreas.schaefer@physik.uni-regensburg.de}

\maketitle

\begin{history}
%\received{(received date)}
%\revised{(revised date)}
%\accepted{(Day Month Year)}
%\comby{(xxxxxxxxxx)}
\end{history}

\begin{abstract} 
We review current ideas on entropy production during the different stages of a relativistic nuclear collision. This includes recent results on decoherence entropy and the entropy  produced during the hydrodynamic phase by viscous effects. We start by a discussion of decoherence caused by gluon bremsstrahlung in the very first interactions of gluons from the colliding nuclei.
We then present a general framework, based on the Husimi distribution function, for the calculation of entropy growth in quantum field theories, 
which is applicable to the early (``glasma'') phase of the collision during which most of the entropy is generated. 
The entropy calculated from the Husimi distribution exhibits linear growth when the quantum field contains unstable modes and the growth rate is asymptotically equal to the Kolmogorov-Sina\"i (KS) entropy. 
We outline how the approach can be used to investigate the problem of entropy production in a relativistic heavy-ion reaction from first principles. 
We show that the same result can be obtained in the framework of a completely different approach called {\em eigenstate thermalization hypothesis}. 
Finally we discuss some recent results on entropy production in the strong coupling limit, as obtained from AdS/CFT duality.  
\end{abstract} 

\section{Overview}\label{sec:intro}
\subsection{The General Problem}

High-energy heavy-ion collisions show a vary rapid transition from a quantum mechanical initial state to a thermalized state known as a quark-gluon plasma. This poses a fundamental, conceptual theory problem, namely how the necessary production of entropy can be reconciled with the T-invariance of quantum chromodynamics (QCD).  One could argue in analogy to the situation in the Einstein-Rosen-Podolsky paradox that all produced particles stay in a highly entangled state which only collapses when a measurement is performed and that entropy production occurs only in that moment.  Nevertheless, a hydrodynamic description of the quark-gluon plasma seems to be possible long before any measurement takes place. The resolution of this apparent contradictionis of interest for a wide spectrum of physical phenomena, e.~g., reheating of the early universe after the end of the presumed period of inflationary growth, when the universe transitioned from an exponentially inflated vacuum bubble to the hot cosmos we observe today. Another example is the gravitational collapse of matter into a black hole, which behaves as if it were a thermal object. In this review we will focus on the thermalization of strongly interacting matter after the collision of two relativistic heavy nuclei, thereby forming a quark-gluon plasma, a process which has the immense advantage that it can be studied in great detail in laboratory experiments. For example it is known that the quark gluon matter in a heavy-ion collision behaves like a nearly ideal thermal fluid already after $1-2$ fm/$c$.

Justifying the use of thermodynamics in the context of quantum field theory is an extremely difficult task in any case, even for phenomena described by quantum electrodynamics (QED), where the quantum field theory is under excellent control. For QCD the situation is even more difficult due to the strong coupling dynamics, which permits analytical calculations only in very much constrained situations. To illustrate the extent of the problem let us note that even in quantum mechanics the problem of thermalization is far from being settled. A nice example is a recent study\cite{Rigol:2009zz}, where a finite one-dimensional quantum system was shown to thermalize only when the so called ``eigenstate thermalization hypothesis'' (ETH\cite{Srednicki:1994xx}) is satisfied. Rigol\cite{Rigol:2008fk} stresses that quantum and classical thermal states have very different natures. ETH claims that each eigenstate of the Hamiltonian is related to a thermal state which is reached by the ``de-phasing'' of the coherent components of the quantum state. 

While we do not want to discuss the merits of this specific concept in more detail, it serves as an example illustrating the fact that the de-phasing or decoherence of a quantum state constitutes one way (if not {\em the} way) to generate entropy. Generated entropy generally is proportional to the number of lost bits of information.  Consequently phases carry an important part of the information contained in a quantum state, their loss generates a significant amount of entropy. 

A general mechanism for losing information about the state of a quantum system is the entanglement of its wave function with that of its environment. This mechanism obviously does not apply to an isolated system.  The conservation of the von Neumann entropy for an isolated system implies that any information encoded in its quantum state is never lost, at least in principle.  However, the information may become encoded in observables of such complexity or requiring such a degree of experimental precision that its retrieval is impractical. A standard approach to describe this effective loss of information in an isolated system is {\em coarse graining}. By averaging over finite regions of phase space, information is lost, and thus entropy is produced. 

Often, coarse graining is discussed in connection with measurements performed on a system. However, in Section 2 we will discuss how even in the absence of physical measurements the uncertainty principle implies a natural, measurement independent coarse graining. It thus implies effective entropy production for nonlinear, time-reversal invariant systems, in which individual trajectories approach all points in some region of phase space, while all ensembles of trajectories respect Liouville's theorem, i.e. occupy a constant phase space volume.

This argument also applies to the information residing in the phases between the different components of a quantum system, which are smeared by coarse graining. Thus decoherence or de-phasing plays a crucial role for the discussion of thermalization and entropy production, which naturally leads to the question whether decoherence alone is sufficient to reach thermalization. This is not the case, because decoherence may lead to systems with statistical features, which have energy and momentum distributions that are far from equilibrium. 

\subsection{Relativistic Heavy Ion Collisions}

Heavy-ion collisions provide a good example for such systems. After decoherence of the initial parton distributions in the colliding nuclei one is left with some sort of quark-gluon matter that can, for most part, be treated statistically and semi-classically. The natural question is whether the equations of fluid dynamics can be applied to describe the subsequent evolution of this matter. Fluid dynamics is an effective theory for the time evolution of the expectation value of the components of the energy-momentum tensor $T_{\mu\nu}$, which encodes the dynamics of the system in terms of macroscopic quantities: energy density, pressure, collective flow velocity, and the equation of state.  Its applicability requires that spatial and temporal gradients of these quantities are small, and that the deviations of the microscopic statistical state of the system from local thermal equilibrium are small and can be represented by viscous (gradient) corrections describing anisotropies of the stress tensor. 

The early-time energy momentum tensor for heavy-ion collisions (colliding along the $z$ axis) obtained by solving the classical Yang-Mills equations of motion has the approximate form\cite{Fries:2007iy}
\begin{equation}
T_{\mu\nu} \approx \left(
\begin{array}{cccc}
\epsilon &0&0&0\\
0 & p &0&0 \\
0& 0& p& 0 \\
0& 0& 0& -p\cr
\end{array}
\right) 
\end{equation}
with negative longitudinal pressure, which makes it unsuitable for hydrodynamical calculations. The negative longitudinal pressure is an expression of the fact that Gauss' law does not permit all components of the gauge field to decohere, because they are not independent of the field generating color charges. In this case, secondary gluon production depletes the remaining coherent longitudinal fields and eventually generates positive longitudinal pressure. Only then can the system establish acceptable initial conditions for the following hydrodynamical evolution. The particle production process and the interaction among the produced particles therefore must be included in the description of the thermalization process. 

A natural question to ask in this situation is, how quickly the components of the stress tensor develop a sufficient degree of isotropy in the locally co-moving frame for a hydrodynamical description to be valid, and how this {\em isotropization time}, $\tau_{\rm iso}$, compares to the {\em decoherence time}, $\tau_{\rm dec}$.   Note that isotropization is a less stringent requirement than thermal equilibration. In principle, 
an isotropic stress tensor could be generated by an isotropic, yet strongly non-thermal gluon distribution.  Indeed, if one considers a dilute ensemble of de-phased particles, which is described by the Boltzmann equation, in a state near thermal equilibrium, its relaxation to equilibrium is governed by an infinite set of relaxation times and their associated modes. Because of the local conservation of energy and momentum, the viscous modes are usually those which survive at late times, and thus entropy growth at late time is usually controlled by the viscosities and the thermal conductivity. But this does not have to be the case when the system is far off equilibrium, where unstable, exponentially growing modes may exist that do not permit a linearized description. Such processes can somtimes be interpreted as run-away (stimulated) particle production.

While particle production is usually not interpreted as decoherence, it may in fact be related to it. For example, bremsstrahlung can be understood as decoherence of the coherent, quasi-real Weizs\"acker-Williams fields associated with any fast moving charge. While such radiation does not result in exponential growth in QED in the vacuum, it can do so in the presence of strong fields (e.g.\ in a free electron laser) or in the nonlinear dynamics of an electromagnetic plasma or a quark-gluon plasma far off equilibrium. The interpretation of how entropy is generated in such processes may then depend on the theoretical and conceptual framework in which the processes that lead to thermalization are described.

\subsection{General Considerations}

The question of whether, how, and how fast a highly excited state of QCD matter equilibrates thermally is thus a multi-faceted one, and we certainly do not profess to fully understand all relevant points of view. Below, we will outline and discuss those we are aware of, without claiming completeness. A general framework for the study of thermalization processes starts from the observation that thermal equilibrium corresponds to the state of highest entropy. Our approach thus focuses on the rate at which entropy is produced in the system. As already mentioned, the von Neumann entropy of an isolated quantum system never increases, because the time evolution of an isolated quantum state is unitary. However, when we speak of the thermalization of an isolated system, we mean that its state ``looks'' thermal to most practically feasible experiments that measure a select number of observables. Owing to the lack of complete observation we replace the microscopic, possibly pure quantum state of the system by a density matrix that traces over all unmeasured observables. By this coarse graining procedure we assign an entropy to the system, which can grow as a function of time if the microscopic structure of the quantum state gets hidden in more and more complex, and hence practically unobservable, degrees of freedom. The coarse grained, apparent entropy is then a measure of the complexity of the microscopic state of the system.  

\begin{figure}[ht]
\centerline{
\includegraphics[width=0.7\textwidth]{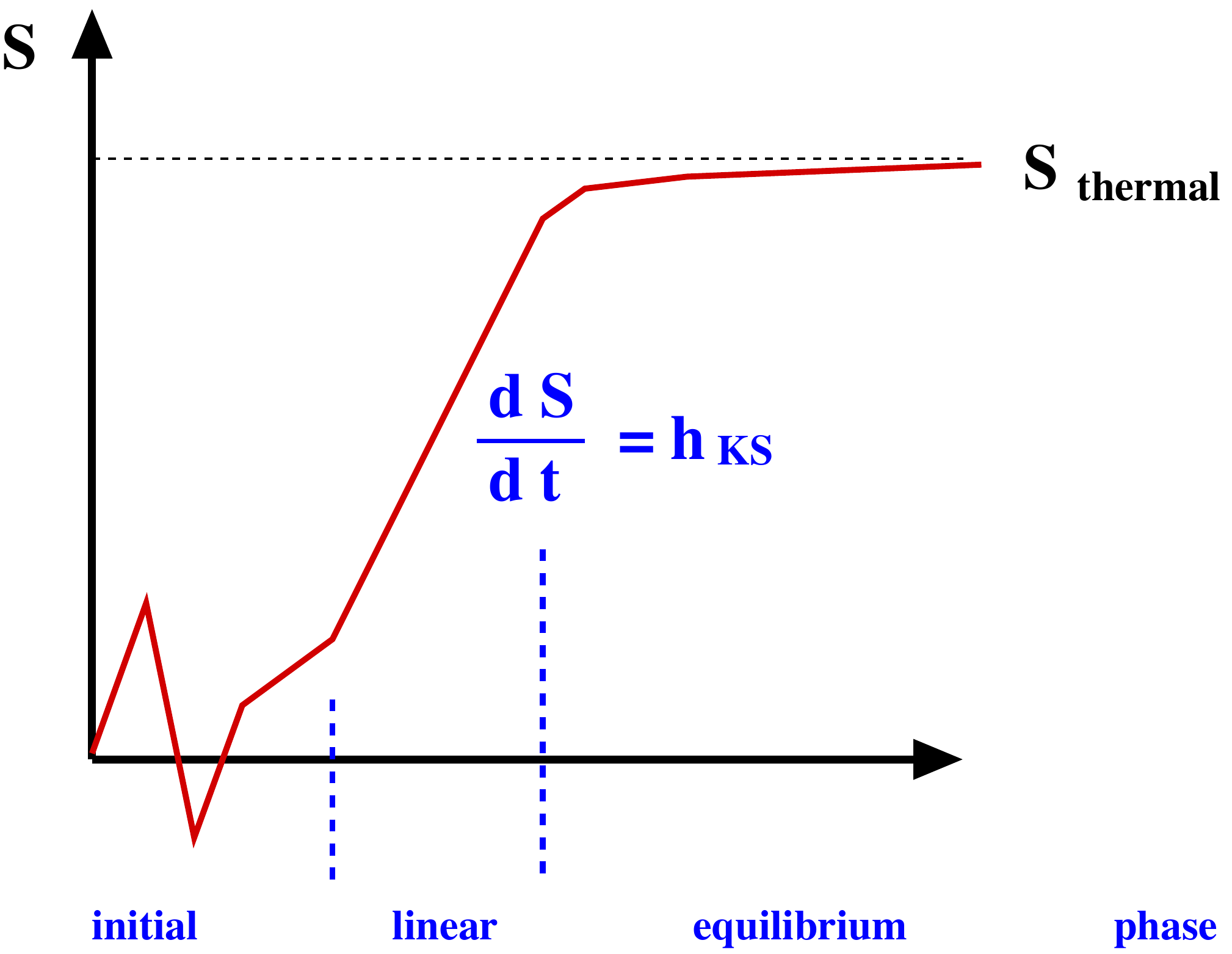}}
\caption{Sketch of entropy growth in a generic non-linear system. Initially the behavior is non-universal, depending on the specific initial conditions, then one encounters a phase of linear entropy growth and finally an asymptotic approach of complete thermal equilibrium }
\label{fig:AS1}
\end{figure}

The generic time evolution of the coarse grained entropy for a dynamical system with ergodic properties is sketched in Fig.\ref{fig:AS1}. At very early times, the behavior is extremely sensitive to details of the initial state and will depend on the initial phase correlations and the occupation probabilities of the various eigenmodes of the system. If all modes are randomly occupied the length of this period can be estimated\cite{Kunihiro:2010tg} to be of order $\sqrt{N}$, where $N$ is the number of independent modes. The period of approximately linear rise of the coarse grained entropy is governed by the sum of all positive Lyapunov exponents of the system in the classical limit, the so-called Kolmogorov-Sina\"i (KS) entropy growth rate. The rise starts when the amplitude of the unstable modes has outgrown the average mode occupation, and it terminates when the coarse grained entropy approaches the equilibrium entropy. While the rate of growth of the entropy is an intrinsic property of the system, the total duration of the quasi-linear period depends 
on the specific nature of the initial state. Finally, the last phase is governed by the relaxation times of small deviations from equilibrium.

The dynamical evolution of the quark-gluon plasma created in a relativistic heavy ion collision is commonly described by relativistic viscous hydrodynamics, which is well suited to describe the last, asymptotic phase sketched in Fig.\ref{fig:AS1}. Obviously, the question, when the transition from the microscopic description to the hydrodynamical one should be taken, has no unique answer but has to be decided on the basis of heuristic and practical considerations. In principle, the initial conditions for the hydrodynamic simulation will depend on which moment is chosen as the starting time.  However, if hydrodynamics applies at all, the late time results of the simulation will depend only little 
on when exactly the transition from a microscopic description to hydrodynamics is made.  According to our discussion, the only constraint is that this time should be chosen close to the end of the phase of linear entropy growth.

In fact, the situation is even more involved. In practice, the information about the supposedly thermal character of the final state is obtained from single-particle observables, including emitted particle yields, spectra, and angular distributions, as well as of selected two-particle correlation functions, such as balance functions.  These observables are all of hadronic nature; thus, one also has to take into account the entropy production which occurs during hadronization, as well as that caused by interactions in the expanding hadron gas before freeze-out. These final one-particle or limited two-particle observables trace over a very large number of degrees of freedom and thus imply a highly coarse grained description in terms of the fundamental degrees of freedom. However, we are not so much interested in this late state as in the question when the transition to a hydrodynamical description is justified early in the reaction. For this question, a coarse grained entropy definition that is dictated by the exigencies of final-state measurements, long after the collision is over, makes little sense, because a thus defined entropy may have very little relevance for the dynamics governing the time evolution of the system.  This suggests that it may be useful to study entropy growth under a {\em minimal} amount of coarse graining (as imposed by the uncertainty principle, see below) and to use this approach to estimate the rate at which the coarse grained entropy approaches its thermal equilibrium value.

The different conceptual approaches to the thermalization problem are to some extent complementary. Obviously, when the system truly equilibrates, all approaches will indicate thermalization. However, isotropy of its macroscopic properties alone is a poor measure of thermal equilibrium of a quantum system, nor does the growth of the coarse grained entropy ensure that the system approaches local isotropy. A fluid dynamical description with a thermal equation of state requires both. In this review, we focus on the progress of our understanding of how and when the apparent entropy in the final state of a relativistic heavy ion collision is generated by the internal dynamics of the highly excited QCD matter. However, we caution the reader that, while substantial progress has been made over the last years, we are still far from having a complete answer to the problem of thermalization.

\section{Entropic History of a Relativistic Heavy Ion Collisions}\label{sec:history}

\subsection{Stages of entropy production}

As already mentioned, much of our information on the behavior and properties of hot QCD matter is derived from measurements of the particle yields and spectra in the final state of relativistic heavy ion collisions and their interpretation in terms of thermodynamic and hydrodynamic concepts. At the energies available at the Relativistic Heavy Ion Collider (RHIC) and the Large Hadron Collider (LHC) one of the most relevant quantities is the azimuthal quadrupole anisotropy of the collective flow, usually called {\em elliptic flow} and denoted as $v_2$. An excellent agreement of hydrodynamical calculations of the flow anisotropy of the matter produced in the nuclear collisions with the elliptic flow measurements requires the assumption of a rapid thermal equilibration of the matter on a time scale of the order of 1 fm/c. An important problem in the description of relativistic heavy ion reactions is thus to understand how the produced matter equilibrates so quickly, {\em i.e.}, when and how entropy is created in the reaction.  In principle, one can distinguish five different stages of entropy production:
\begin{enumerate}
\setlength{\itemsep}{0pt}
\item Decoherence of the initial nuclear wave functions;
\item Thermalization of the partonic plasma (``glasma'');
\item Dissipation due to shear viscosity in the hydrodynamical expansion;
\item Hadronization accompanied by large bulk viscosity;
\item Viscous hadronic freeze-out.
\end{enumerate}

\begin{figure}[hbt]
\centerline{\includegraphics[width=0.85\textwidth]{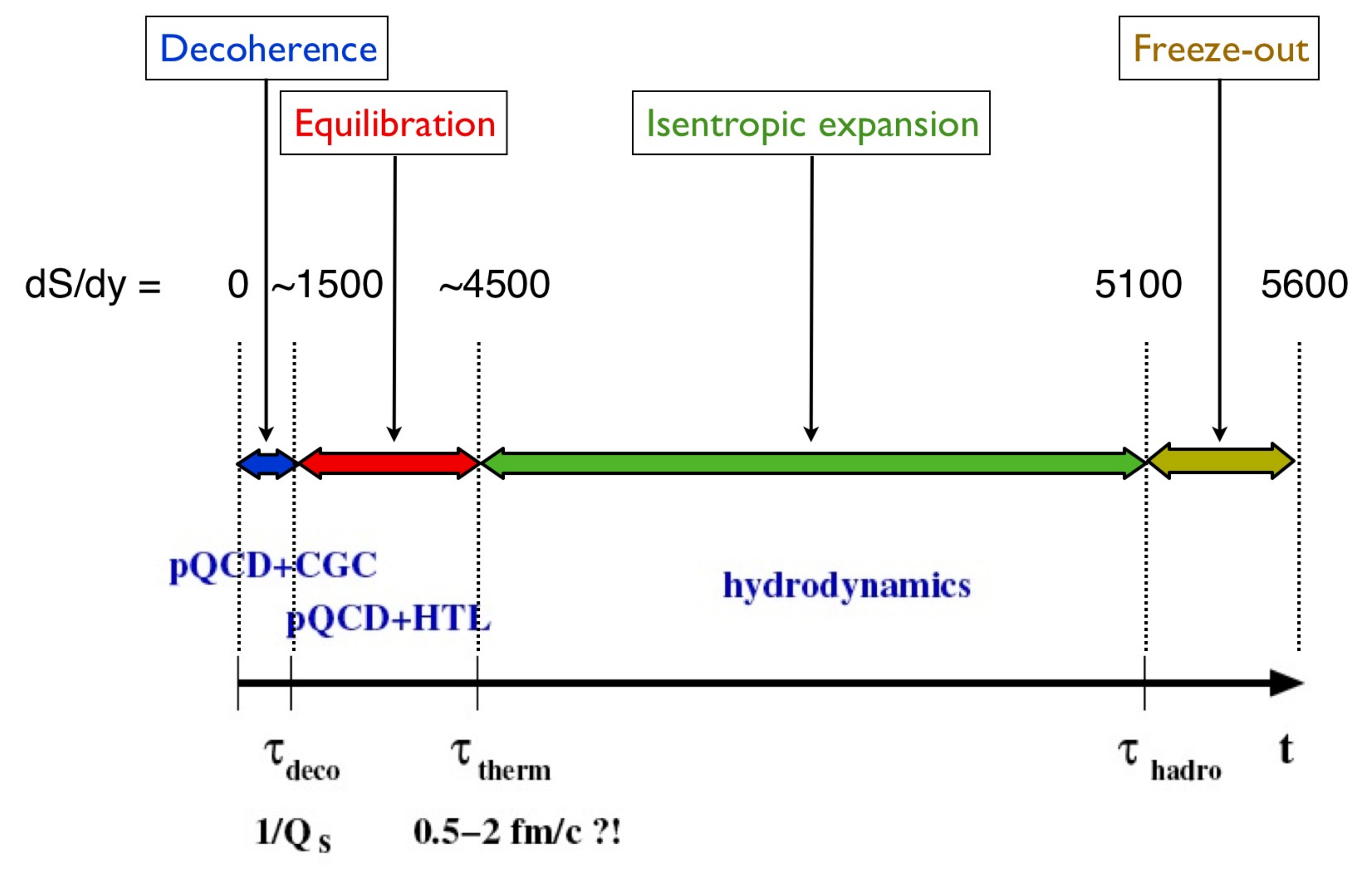}}
\caption{Entropic history of a central Au+Au collision at top RHIC energy. The values of $dS/dy$ indicate the entropy per unit rapidity reached at the end of various collision stages based on experimental data and model estimates.}
\label{fig:BM1}
\end{figure}

Our present incomplete knowledge about the contribution of these different stages to the final entropy is depicted in Fig.~\ref{fig:BM1}. Before we discuss these stages in turn, we note that the final entropy per unit rapidity $dS/dy$ is one of the best known quantities in relativistic heavy ion physics. In the case of Au+Au collisions at RHIC the entropy distribution $dS/dy$ at particle freeze-out can be determined from an analysis of the final hadron spectra in combination with the information on the source radius derived from identical particle (HBT) correlations.\cite{Pal:2003rz}\ 
The slightly extrapolated result for the 6\% most central Au+Au collisions at $\sqrt{s_{\rm NN}}= 200$ GeV is $(dS/dy)_{\rm f}=5600 \pm 500$ at midrapidity.\cite{Muller:2005en}\ Alternatively, the final entropy can be deduced from the measured hadron abundances, combined with the calculated entropy per particle for a hadron gas in chemical equilibrium at $T_c\approx 160$ MeV, $S/N\approx 7.25$, which yields the result\cite{Muller:2005en}\ $(dS/dy)_{\rm ch}=5100 \pm 200$. The 10 percent difference between these values can be attributed to the entropy production during the hadronic freeze-out and reflects the entropy production due to the decay of excited hadronic resonance states as well as the contribution from the substantial shear viscosity of a thermal hadron gas.\cite{Demir:2008tr}

\subsection{Decoherence}

Entropy production by decoherence dominates the very first stage of a heavy ion collision. The loss of coherence is measured by the decay of off-diagonal elements of the density matrix $\rho$ describing the system. A practical way of investigating this process is to calculate the decay rate of the quantity ${\rm Tr}\rho^2/[{\rm Tr}\rho]^2$. The time scale of the decoherence of the initial nuclear wave function has been studied by this method in the color glass condensate model (CGC).\cite{Muller:2005yu,Fries:2008vp}\ In this model, the initial nuclear gluon distributions are coherent over a wide range of longitudinal momenta $k_L$ (in the beam direction) due to the high degree of Lorentz contraction of the fast moving nuclei. The elementary process driving the decay of the off-diagonal elements of the density matrix is the fusion of one gluon from each nucleus, which removes gluons from the initial-state wave functions. The relevant time scale in this process is set by the saturation scale $Q_s$ of the nuclear gluon distribution. $Q_s$ is a measure of the average transverse momentum of gluons in a fast moving nucleus. Gluons with transverse momentum $k_T \leq Q_s \sim 1-2$ GeV can be thought of as components of a quasi-classical gauge field.  A careful analysis yields the result that the characteristic decoherence time is $\tau_{\rm dec} = c Q_s^{-1}$, where $c$ denotes a calculable constant close to unity.\cite{Muller:2005yu,Fries:2008vp}\ While this result is expected on dimensional grounds, the fact that $c = O(1)$ is important because it shows that  decoherence occurs over a time of less than 0.2 fm/c. 

Here we do not want to present the complete chain of arguments leading to this result, but only provide a few details showing how the CGC model enters into the calculation.  We study perturbative gluon interactions between the gluons in a nucleus at rest (nucleus 1) and a fast moving one (nucleus 2). The time evolution of the density matrix describing nucleus 1 reads
\cite{Lappi:2006fp} 
\begin{equation}
  \rho_{\hat A,A} (t) = \sum_{\hat B,B} \int_0^t d\hat t dt' 
  ~ H_{\hat A,\hat B}^{QCD}(\hat t) \rho_{\hat B,B}(0)H_{B,A}^{QCD}(t')
\end{equation}
The CGC model enters by specifying the correlator $\langle A_\mu^c (p) {A^{\hat c}_{\hat \mu}}^\dagger (q) \rangle_2$ between gauge fields in nucleus 2 as
\begin{equation}
 \left\langle A_\mu^c (p) {A^{\hat c}_{\hat \mu}}^\dagger (q) \right\rangle_2 
= \delta^{c\hat c}  (2\pi)^2 
  \delta^2(\mathbf{p}_\perp - \mathbf{q}_\perp) 
   \frac{\pi^2\delta(p^-)\delta(q^-)}{p^+q^+} 
   \frac{p_{\mu} p_{\hat\mu}}{p_\perp^2} F(\mathbf{p}_\perp)
\end{equation}
where
\begin{equation}
 F(\mathbf{p}_\perp) = \int d^2x e^{-i \mathbf{p}_{\perp}\cdot\mathbf{x}}\frac{4 (N_c^2-1)}{N_c g^2 \mathbf{x}^2} 
  \left( 1- e^{ -g^4 N_c/(8\pi)\mu^2 \mathbf{x}^2 \ln 1/( |\mathbf{x}|\Lambda)} \right) ~~.
\end{equation}
This form implies decoherence in transverse momentum but coherence in longitudinal momentum.  For the gluon field correlator in nucleus 1 (at rest) one makes a simple Gaussian ansatz 
\begin{equation}
  \left\langle {A}^{a}_{k,\lambda} {{A}^{\hat a}_{\hat k,\hat\lambda}}^\dagger \right\rangle_1 
  = \delta_{\hat\lambda \lambda}
   \delta_{\hat{a} a} (2\pi)^4 \delta^4(\hat k - k) \mathcal{N} \zeta^2 
   e^{-\zeta^2({k^0}^2+\mathbf{k}^2)} ~~.
\end{equation}
It turns out that a finite result is only obtained if the following running of the QCD coupling constant is used\cite{Kovchegov:2007vf}
\begin{equation}
g^4 ~\to g^2(\Lambda^2) g^2(1/x^2) ~~,
\end{equation} 
where $\Lambda$ is the infrared cutoff of the CGC, which cancels in the ratio ${\rm Tr}\rho^2/[{\rm Tr}\rho]^2$. This ratio then turns out to be inversely proportional to the life-time of the system, and one can define $\tau_{\rm dec}$ as the time after which this ratio has dropped by a factor $1/e$. 

The amount of entropy created from the loss of coherence of the components of the nuclear gluon wave functions can be estimated as follows. The complete decoherence of a coherent quantum state with average occupancy $\bar{n}$ results in a mixed state with entropy\cite{Muller:2003cr}\ $S_{\rm dec} \approx (\ln 2\pi\bar{n} + 1)/2$. The decoherence occurs in transverse domains of spatial size $Q_s^{-1}$, which are causally disconnected during the decoherence process. The number of causally disconnected transverse domains in a central Au+Au collision is of order $(Q_s R)^{-2}$, where $R$ denotes the nuclear radius. Accounting for a longitudinal coherence length $\Delta y \approx 1/\alpha_s$ and using $\bar{n}\approx 1/\alpha_s$ one finds:\cite{Fries:2008vp,Muller:2003cr}
\begin{equation}
(dS/dy)_{\rm dec} \approx \frac{1}{2} Q_s^2 R^2 \alpha_s (\ln 2\pi/\alpha_s +1) \approx 1,500
\label{eq:dSdy}
\end{equation}
at midrapidity for a central Au+Au collision. This is roughly one quarter of the observed final entropy per unit rapidity.

\subsection{Viscous flow}

The other stage which is reasonably well understood is the hydrodynamic expansion stage. RHIC data on the elliptic flow of hadrons in non-central collisions indicate that the shear viscosity during this phase is small. The bounds on the shear viscosity $\eta$ compatible with the RHIC data lie characteristically in the range $1 \leq 4\pi\eta/s \leq 2.5$.\cite{Romatschke:2007mq,Gavin:2007zza,Luzum:2008cw,Song:2008hj,Lacey:2009xx,Song:2010mg}\ Such a small viscosity cannot increase the total entropy of the expanding fluid by much. Systematic studies of the entropy growth in longitudinally boost-invariant, viscous fluid dynamics for different initial conditions were performed by Fries {\em et al}.\cite{Fries:2008ts}\ The viscous stress-energy tensor in the locally co-moving frame has the form $T^{\mu\nu} = {\rm diag}(\varepsilon,P_{\perp},P_{\perp},P-z)$. The transverse and longitudinal pressures can be decomposed into the thermodynamic pressure $P$ and the components $\Phi$ and $\Pi$ of the shear and bulk stress: $P_{\perp} = P + \Pi + \frac{1}{2}\Phi$; $P_{z} = P + \Pi - \Phi$.

At early times, $\Phi$ and $\Pi$ are given by the initial conditions for the stress tensor established during the thermalization process.  In view of the dilution effect of the longitudinal expansion imprinted on the matter by the collision, the pressure components are expected to satisfy the ordering $P_{z} < P_{\perp}$. The smallest physically meaningful value of the longitudinal pressure is $P_{z} = - P_{\perp}$, which arises when the matter is completely in the form of coherent longitudinal fields.\cite{Fries:2007iy,Kovchegov:2007pq}\  However, the fluid dynamical approach does not apply to such a situation, because it assumes small deviations from local equilibrium. The minimal condition for fluid dynamics to be applicable is the mechanical stability condition: $P_z, P_\perp \geq 0$, but the regime where the fluid dynamical description can be trusted only begins when $\Phi/P, \Pi/P \ll 1$.

The equations governing the longitudinal expansion of the medium are:\cite{Heinz:2005zi,Muronga:2003ta,Baier:2006um,Baier:2007ix}
\begin{eqnarray}
  \frac{\partial\varepsilon}{\partial\tau} &=& - \frac{1}{\tau}(\varepsilon + P + \Pi - \Phi) \, ,  
  \label{eq:evol1} \\
  \tau_{\pi} \frac{\partial\Phi}{\partial\tau} &=& \frac{4\eta}{3\tau} -  \Phi(\tau) -
  \frac{4\tau_{\pi}}{3\tau}\Phi +\frac{\lambda}{2\eta^2}\Phi^2 \, , 
  \label{eq:evol2} \\
  \tau_{\Pi} \frac{\partial\Pi}{\partial\tau} &=& - \frac{\zeta}{\tau} - \Pi(\tau) .
\label{eq:evol3}
\end{eqnarray}
$\eta$ and $\zeta$ are the shear and bulk viscosity, respectively. The relaxation times for the shear and bulk stress, $\tau_{\pi}$ and $\tau_{\Pi}$, were assumed to be those derived from the Boltzmann equation, $\lambda$ was taken from the supersymmetric gauge theory (see Fries {\em et al.}\cite{Fries:2008ts}\ for details). The entropy per unit rapidity and transverse area, $dS/(dydA) = \tau s$ obeys the equation:\cite{Heinz:2005bw}
\begin{equation}
  \frac{\partial(\tau s)}{\partial\tau} = \frac{\tau}{T} \left( \frac{3 \Phi^2}{4 \eta} + \frac{\Pi^2}{\zeta} \right).
  \label{eq:dsdtau}
\end{equation}

\begin{figure}[tb]   
\centerline{
\includegraphics[width=0.6\linewidth]{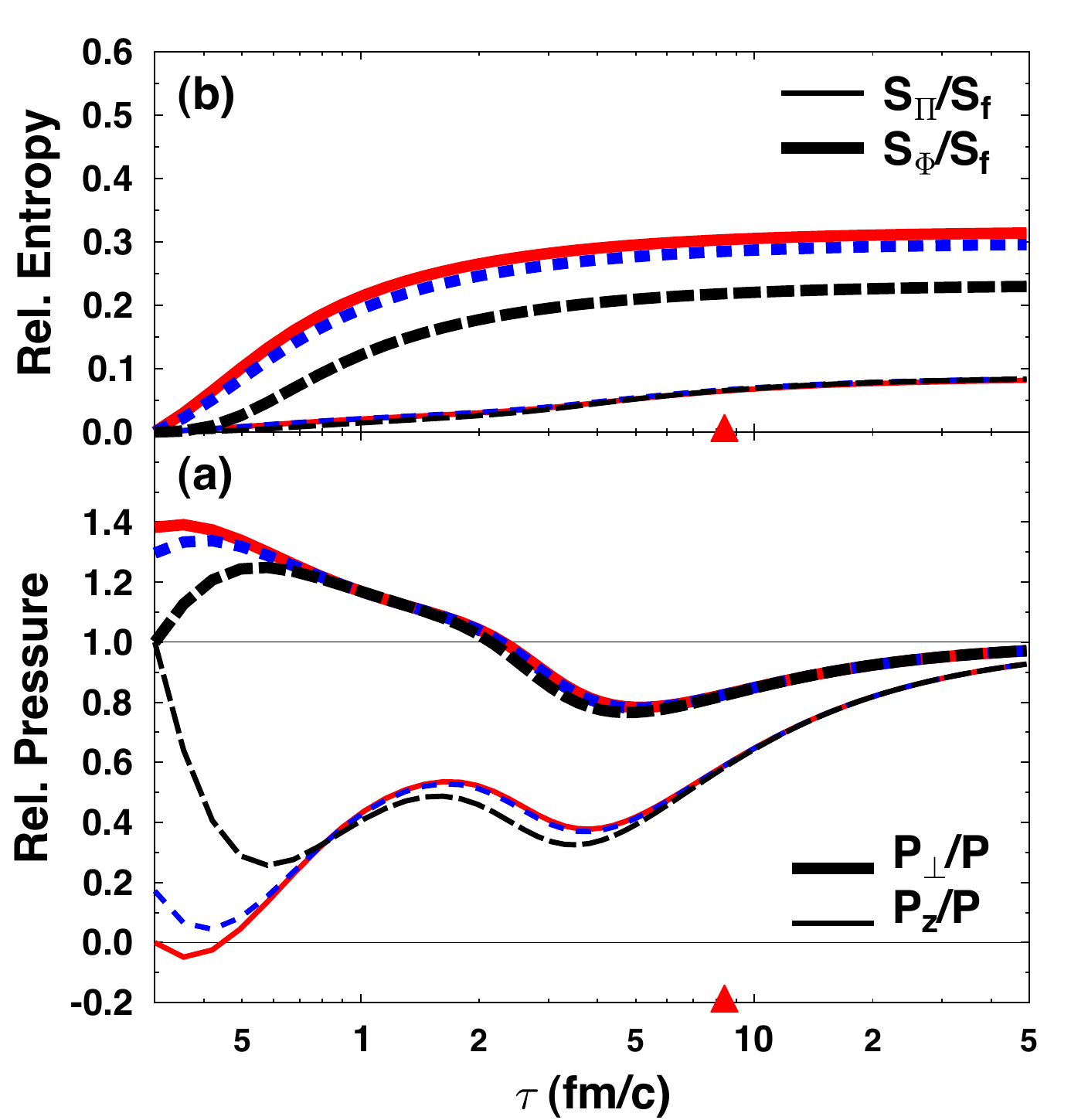}}
\caption{
(a) Relative transverse and longitudinal pressure, $P_{\perp}/P$ (dashed) and $P_{z}/P$ (solid), as functions of time $\tau$ for the initial conditions (i) (black), (ii) (blue), and (iii) (red). 
(b) Relative entropy production from bulk and shear stress, $S_{\Pi}/S_f$ (solid) and $S_{\Phi}/S_f$ (dashed) for the same scenarios. For further details, see Fries {\em et al}.\protect\cite{Fries:2008ts}
}
\label{fig:visc-hydro}
\end{figure}

\begin{figure}[tb]   
\centerline{
\includegraphics[width=0.6\linewidth]{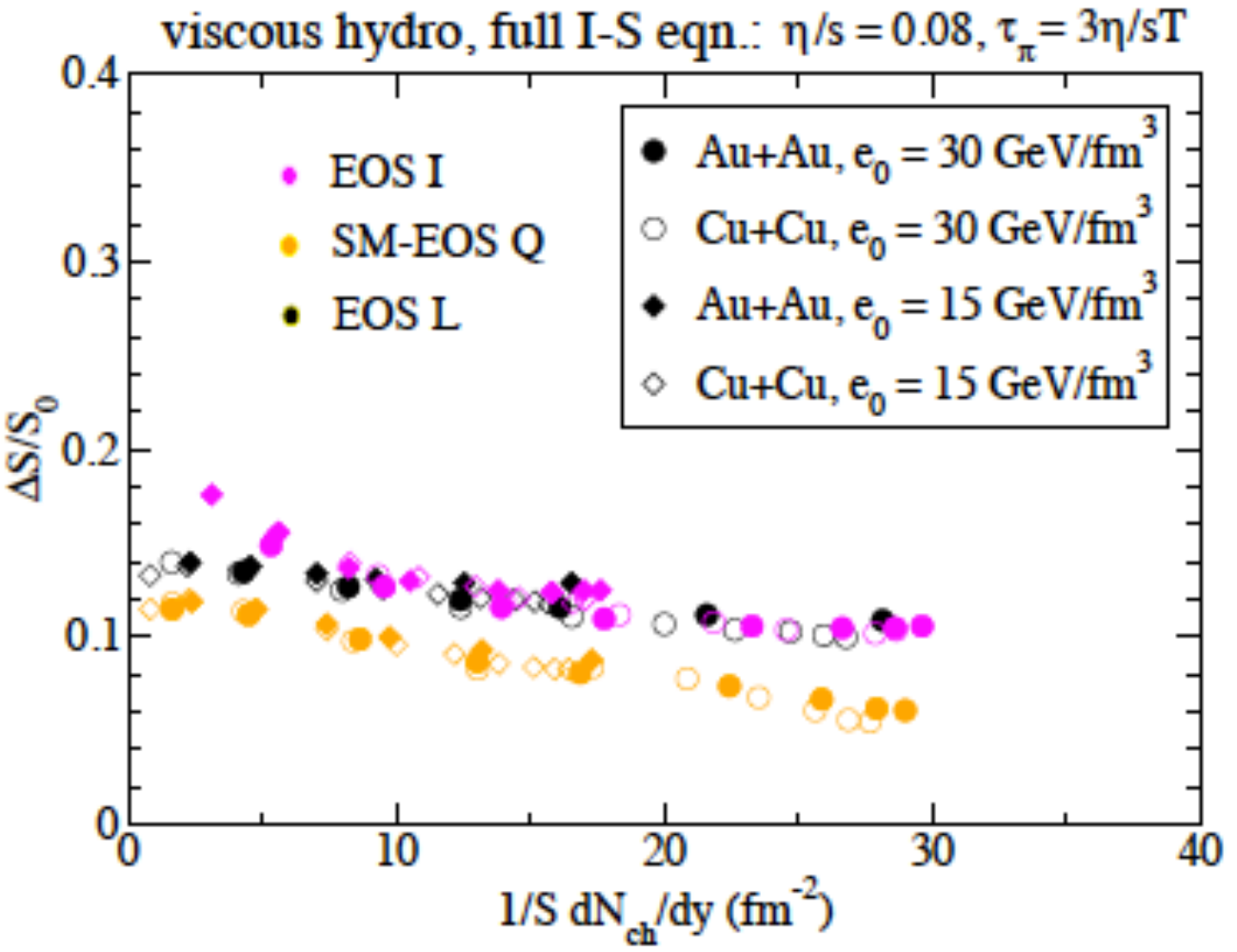}}
\caption{
Relative entropy increase due to bulk and shear stress, $\Delta S/S_0$, for two-dimensional boost invariant viscous hydrodynamical expansion of a quark-gluon plasma for three different equations of state. The entropy increases by approximately 10 percent, if the shear visscosity is equal to the KSS bound $\eta/s = 1/4\pi$. For further details, see Song and Heinz\protect\cite{Song:2008si}.
}
\label{fig:2d-hydro}
\end{figure}

Fries {\em et al.}\cite{Fries:2008ts}\ started the hydrodynamical simulations after an initial time $\tau_0 = 0.3$ fm/$c$, which is compatible with the expected decoherence time but considerably shorter than the equilibration times estimated from the elliptic flow analysis. The initial energy density was fixed at $\varepsilon_0 = \varepsilon(\tau_0) = 50$ GeV/fm$^3$. The early starting time was chosen to explore, in a schematic way, how long it would take to reach the domain of applicability of the hydrodynamical description defined by the conditions
\begin{equation}
  \frac{|P_{\perp}-P|}{P} \leq \frac{1}{2} ; \qquad  \frac{|P_{z}-P|}{P} \leq \frac{1}{2} .
\label{eq:hydro-cond}
\end{equation}
Fries {\em et al.}\cite{Fries:2008ts}\ considered three different initial conditions: 
\begin{itemize}
\item[(i)]  $\Pi(\tau_0) = \Phi(\tau_0) = 0$, corresponding to thermal equilibrium at $\tau_0$;
\item[(ii)] $\Pi(\tau_0) = -\zeta(T_0)/\tau_0$ and $\Phi(\tau_0) = 4\eta(T_0)/(3\tau_0)$, corresponding to the shear stress of the relativistic Navier-Stokes theory;
\item[(iii)] $\Pi(\tau_0) = -\zeta(T_0)/\tau_0$ as in (ii), but $\Phi(\tau_0) = P(\tau_0)+\Pi(\tau_0)$ for vanishing 
initial longitudinal pressure.
\end{itemize}
Typical results from these calculations are shown in Fig~\ref{fig:visc-hydro}. The lower panel (a) indicates that the condition (\ref{eq:hydro-cond}) is approximately satisfied for $\tau > \tau_{\rm eq}\approx 1$ fm/$c$ independently of the choice of the initial condition for the stress tensor. On the other hand, the upper panel (b) shows that most of the viscous entropy production occurs before this time during a period when the validity of the hydrodynamical approach is questionable. The relative contribution to the final entropy from viscous forces after $\tau_{\rm eq}$ does not exceed 10 percent for the parameter choices explored in this calculation. 
The same conclusion was reached by Song and Heinz\cite{Song:2008si}\ in a two-dimensional boost-invariant calculation (see Fig.~\ref{fig:2d-hydro}). We may thus conclude conservatively that the entropy at the moment of local equilibration must be $(dS/dy)_{\rm eq} \ge 4,500$. These considerations tell us that approximately half of the final entropy must be generated during the thermalization process, which cannot be described by fluid dynamics. We next discuss various theoretical approaches to the entropy growth rate during the thermalization stage.

\section{Entropy growth rate}

\subsection{Husimi formalism and KS entropy}

If the quark-gluon plasma were a weakly coupled system, the growth of the entropy during the equilibration phase could be calculated from a partonic Boltzmann equation. However, the small shear viscosity and other observations from the RHIC experiments, such as measurements of jet quenching, indicate that the plasma is strongly coupled. In addition, it is thought that the state created by the decoherence of the nuclear gluon wave functions (the {\em glasma}\cite{Lappi:2006fp,Romatschke:2006nk}) is characterized by strong, still partially coherent color fields. Such fields may even be regenerated by plasma instabilities in the pre-equilibrium quark-gluon plasma.\cite{Mrowczynski:2005ki}\  A formalism that professes to describe entropy production during this early stage must, therefore, be able to describe the growing complexity of a quantum system as it makes the transition from a regime of gauge field dominance to the hydrodynamical regime characterized by complete randomness on thermal length scales. 

A general formalism that can describe this transition in terms of the gauge field dynamics has been proposed by Kunihiro {\em et al}.\cite{Kunihiro:2008gv}\ The idea underlying this approach is that the growing entropy measures the increasing intrinsic complexity of the quantum state of the system after appropriate coarse graining. As mentioned in the Introduction, the problem is how to impose a minimal amount of coarse graining without assuming the answer. A general solution to this problem dates back to Husimi.\cite{Husimi:1940xx}\ The Husimi distribution is defined as a convolution of the Wigner function $W(p,x,t)$ of the system with a minimum-uncertainty Gaussian wave packet:
\begin{equation}
H_\Delta(p,x;t) = \int \frac{dp'\, dx'}{\pi\hbar}
\exp\left( - \frac{1}{\hbar\Delta}(p-p')^2 - \frac{\Delta}{\hbar}(x-x')^2 \right) W(p',x';t).
\label{eq:Husimi}
\end{equation}
Here $x$ and $p$ stand for all ``position'' and ``momentum'' variables characterizing the system, which may include particle positions as well as field amplitudes.\cite{Mrowczynski:1994nf}\  The Husimi distribution depends on the squeezing parameter $\Delta$, which measures the ratio of position and momentum uncertainty. Its value can be chosen at liberty, but once fixed, will not change with time. The Husimi distribution can be understood as a coarse-grained phase space distribution of the quantum system, where the coarse graining corresponds to the projection on a coherent state. We recall that coherent states are the closest quantum analogues of classical systems compatible with the uncertainty relation. 

Because the Husimi distribution can be shown to be positive (semi-)definite, it permits the definition of a coarse-grained entropy, first introduced by Wehrl:\cite{Wehrl:1978zz}
\begin{equation}
S_{\rm H,\Delta}(t) = - \int \frac{dp\, dx}{2\pi\hbar} H_\Delta(p,x;t) \ln H_\Delta(p,x;t).
\label{eq:SH}
\end{equation}
Quantum systems containing unstable modes, {\em i.e.}\ modes with an exponentially growing amplitude, have a linearly growing Husimi-Wehrl entropy.\cite{Kunihiro:2008gv}\ The entropy growth rate is found to be independent of the squeezing parameter $\Delta$ and thus independent of details of the coarse graining. It is given by the sum of the exponential growth rates of all unstable modes. In classical dynamical systems, this quantity is known as the Kolmogorov-Sina\"i entropy, or KS entropy, and defined as the sum over all positive Lyapunov exponents $\lambda_k$ of the system:
\begin{equation}
dS_{\rm H,\Delta}/dt \longrightarrow S_{\rm KS} = \sum_{k} \lambda_k \, \theta(\lambda_k).
\end{equation}
The KS entropy is understood to be a measure of the growth rate of the coarse grained entropy of a dynamical system starting from a configuration far away from equilibrium, after an initial start-up phase during which unstable fluctuations grow to dominance and before it gets too close to its micro-canonical equilibrium.\cite{Latora:1999xx}

\begin{figure}[h]
\centerline{
\includegraphics[width=0.46\linewidth]{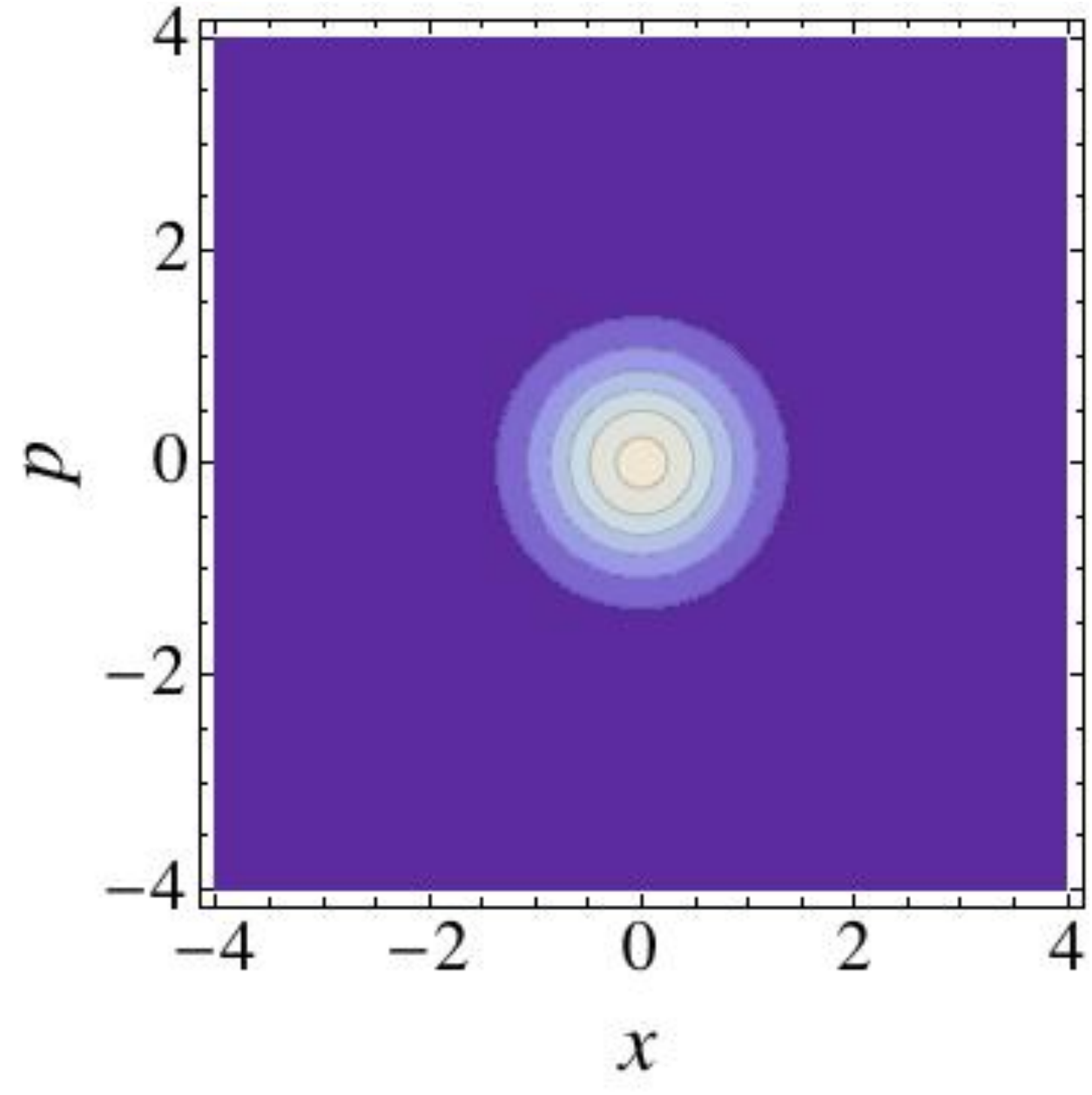}
\hspace{0.04\linewidth}
\includegraphics[width=0.46\linewidth]{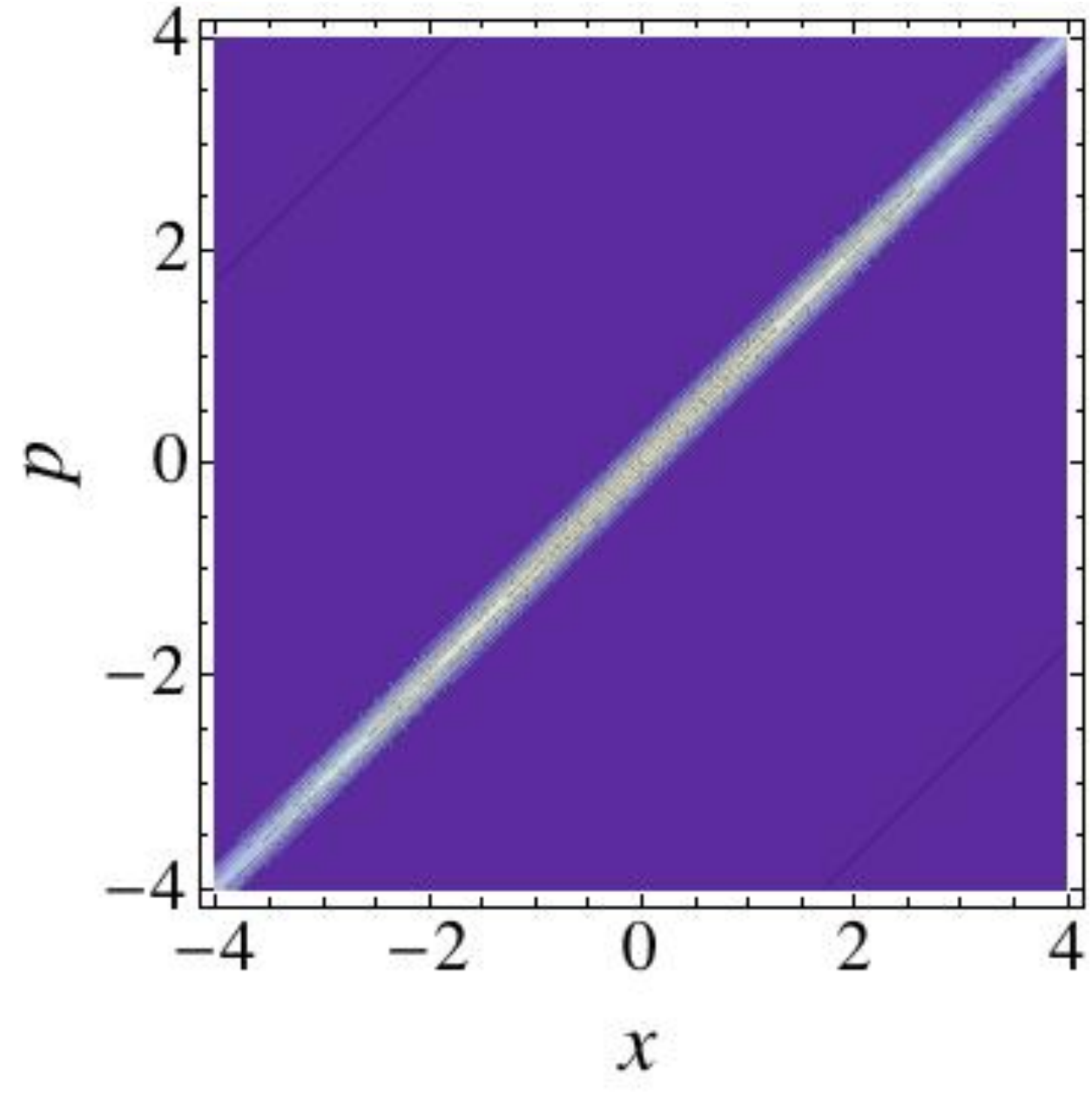}}
\caption{The Wigner function for the inverted oscillator at $t=0$ and $t=2/\lambda$. 
The horizontal axis denotes the scaled position $\omega x$; the vertical axis represents the 
momentum $p$.}
\label{Fig:husimi_1}
\end{figure}
\begin{figure}[h]
\centerline{
\includegraphics[width=0.46\linewidth]{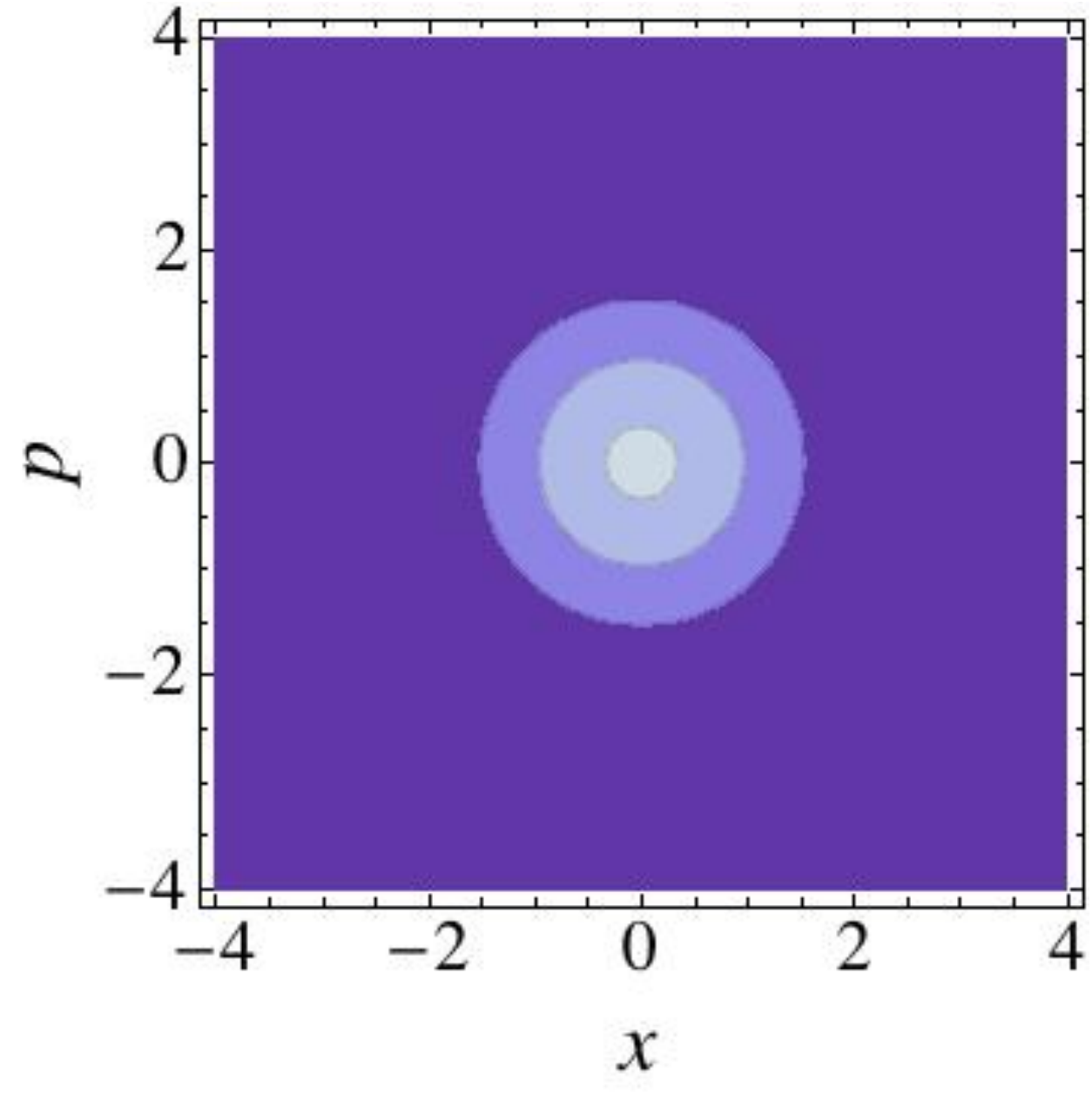}
\hspace{0.04\linewidth}
\includegraphics[width=0.46\linewidth]{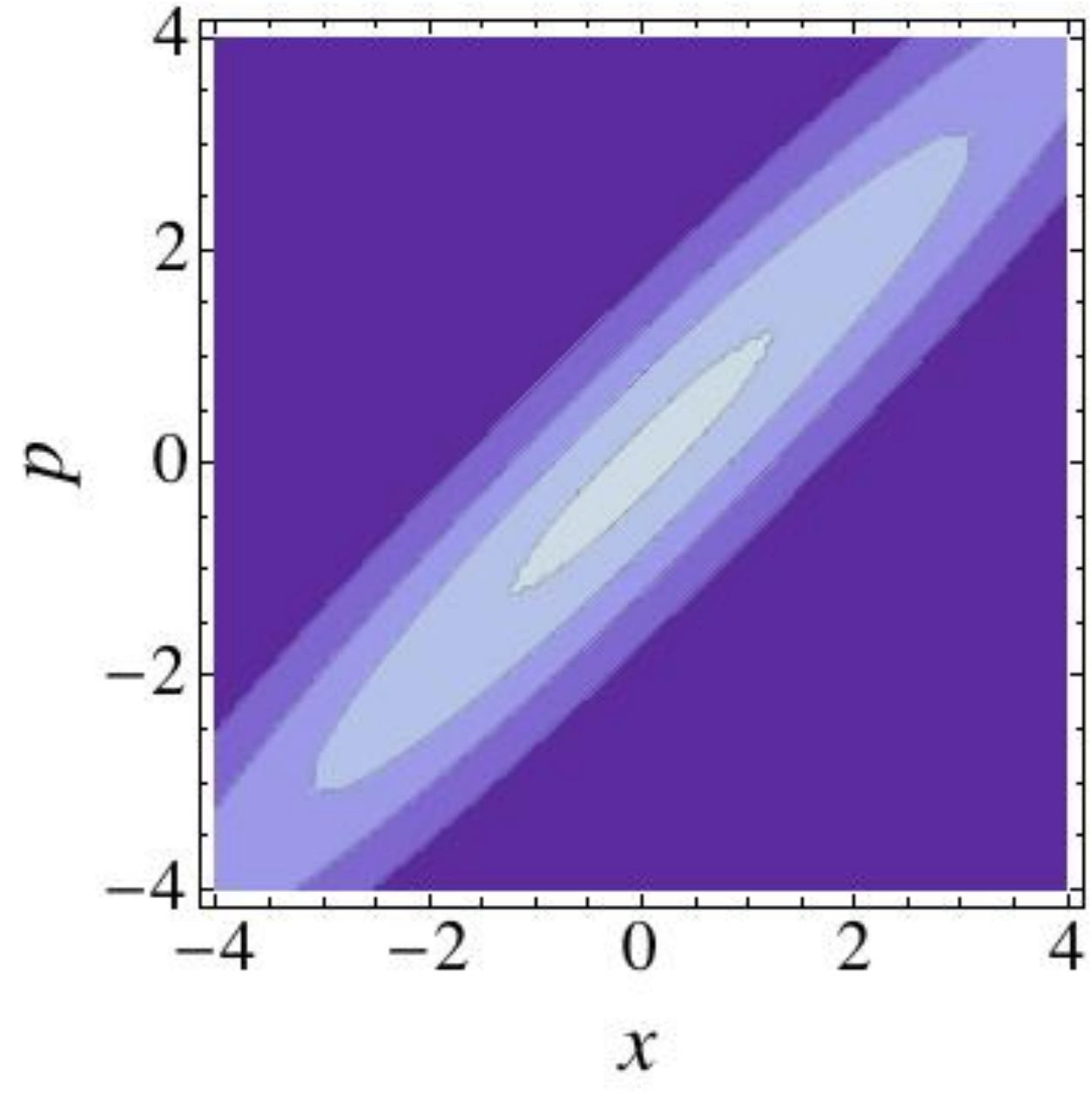}}
\caption{Husimi function \eqref{eq:Husimi} for the unstable oscillator at $t=0$ and $t=2/\lambda$. 
Note that the extent of the distribution in the off-diagonal direction ($p-\omega x$) does not 
shrink beyond the resolution limit set by the Gaussian smearing introduced by the Husimi transform.}
\label{Fig:husimi_2}
\end{figure}

\subsection{Eigenstate Thermalization}

As stated in the Introduction, the eigenstate thermalization hypothesis\cite{Deutsch:1991xx,Srednicki:1994xx} (ETH) states that an isolated quantum system thermalizes under its own internal dynamics if every energy eigenstate contains a thermal component. More precisely, the ETH posits\cite{Rigol:2008fk}\ that the expectation	value $\langle\Psi_E | A | \Psi_E\rangle$ of a few-body observable $A$ in an energy eigenstate $\Psi_E$ of the Hamiltonian of a large, interacting many-body system equals the microcanonical average of $A$ at the mean energy E:
\begin{equation}
\langle\Psi_E | A | \Psi_E\rangle = \langle A \rangle_{E,\rm mc} ~~.
\end{equation}
For a system with very many degrees of freedom, the microcanonical average is generally agrees with the thermal average for an appropriate choice of the temperature. A corrollary of this hypothesis is that a state that is an eigenstate of a few-body observable $A$ will be a superposition of energy eigenstates
\begin{equation}
\Psi_A = \sum_E C_e \Psi_E
\end{equation}
with probabilities $|C_E|^2$ that thermally distributed: $|C_E|^2 \sim \exp(-E/T)$. As the state evolves with time,
\begin{equation}
\Psi_A(t) = \sum_E C_E e^{-iEt/\hbar} \Psi_E
\end{equation}
the phases of the different energy components increasingly diverge from each other. The contributions from different energy eigenstates thus effectively decohere at late times. Any measurement of a few-body observable on the pure quantum state yields a result that is indistinguishable from the results of a measurement on a mixed state defined by the diagonal density matrix\cite{Polkovnikov:2011xx}
\begin{equation}
\rho_{E,E'} = |C_E|^2 \delta_{E,E'} ~~.
\end{equation}
Rigol {\em et al.}\cite{Rigol:2011xx}\ argued on the basis of this argument that the coarse grained entropy of a pure quantum state immediately after a quench can be calculated as the entropy associated with the diagonal energy density matrix in the Hamiltonian after the quench:
\begin{equation}
S = - \sum_E |C_E|^2 \ln(|C_E|^2) ~~.
\label{eq:Sdiag}
\end{equation}
However, this argument does not take into account the time evolution of the phases of the different energy components, which only become quasi-random numbers at late times. At a time $t$ after the quench, the phases of energy eigenstates within an energy band $\Delta E(t) \approx \hbar/t$ differ by less than $\pi/2$, and thus these components of the wave function remain approximately coherent. The time evolution of the coarse grained entropy should therefore be approximately given by the following expression:
\begin{equation}
S(t) = - \sum_{E_\alpha}  |\tilde{C}_{E_\alpha}|^2 \ln(|\tilde{C}_{E_\alpha}|^2) 
\label{eq:St}
\end{equation}
where $E_\alpha$ denotes an energy band of width $\Delta E(t)$ and
\begin{equation}
\tilde{C}_{E_\alpha}(t)  = \sum_{|E-E_\alpha|\leq\Delta E(t)/2} C_E ~~.
\end{equation}
As time progresses, $\Delta E(t)$ continues to shrink, until it is eventually smaller than any spacing between energy eigenvalues of the system. Thus $S(t)$ defined in (\ref{eq:St}) approaches the asymptotic value (\ref{eq:Sdiag}) at late times. However, immediately after the quench, $S(t) = 0$, and the coarse grained entropy only increases gradually as more and more energy eigenstate components dephase from each other. We will show in the next subsection quantitatively how this mechanism operates in the case of an exactly solvable quantum quench.

\subsection{A toy example: The inverted harmonic oscillator}
\subsubsection{Husimi function approach}

The ideas behind the Husimi-Wehrl entropy are best illustrated by a simple example. We choose the inverted harmonic oscillator because this case is so simple that we can analyze it also in a completely different approach built on the {\em eigenstate thermalization hypothesis} (ETH). For the inverted harmonic oscillator 
\begin{equation}
\hat{\cal H} = \frac{1}{2}\hat{p}^2 - \frac{1}{2} \lambda^2 \hat{x}^2 
\end{equation}
we start from a Gaussian wave packet 
\begin{equation}
\langle x | \psi_0 \rangle = \Bigl(\frac{\omega}{\pi\hbar}\Bigr)^{1/4} 
e^{-\omega x^2/2\hbar} 
\label{init}
\end{equation}
and calculate $H_\Delta(p,x;t)$ analytically. The result is shown in Figs.~\ref{Fig:husimi_1} and \ref{Fig:husimi_2}, which illustrate how the phase space volume stays constant for the Wigner function but increases for the Husimi function because the collapse in the shrinking directions is halted by the resolution $\Delta$. One finds that\cite{Kunihiro:2008gv}
\begin{equation}
\lim_{t\rightarrow \infty} \frac{dS_{\rm H,\Delta}}{dt} = \lambda ~~.
\label{eq:dSdt}
\end{equation}
It is noteworthy that this results is independent of $\Delta$ and thus independent of the details of the coarse graining.

\subsubsection{Eigenstate thermalization approach}

Next we show that on gets the same result in the approach based on the eigenstate thermalization hypothesis.  The basic concept underlying the ETH is that every eigenstate of a Hamiltonian contains a thermal state.  The initial quantum state is a coherent superposition of such eigenstates with the property that this thermal component does not contribute to physical observables.  If coherence is lost this state becomes visible and can dominate at late times
\cite{Rigol:2009zz}.
For the inverted harmonic oscillator we can use the WKB approximation
\begin{equation}
\Psi_E = \sqrt{\frac{2}{\pi\hbar}} ~ \frac{\cos \frac{1}{\hbar} \int_0^x dx'~p_E(x')}{\sqrt{p_E(x)}}  ~~,
\qquad
p_E(x) = \sqrt{2E+\lambda^2x^2}
\end{equation}
For the Gaussian initial state (\ref{init}) one expands the initial state in energy eigenstates $\Psi_E$:
\begin{equation}
\Psi(x,t) = \int dE~ C_E~ e^{-\frac{i}{\hbar} Et}~ \Psi_E(x) ~~.
\end{equation}
The time evolution of a narrow energy band $\Delta E$ within the initial state is given by
\begin{eqnarray}
\psi_{E,\Delta E}(x,t) &=& \int_{E-\Delta E/2}^{E+\Delta E/2}~dE' ~C_{E'} ~e^{-iE't/\hbar}~ \Psi_{E'}(x)
\nonumber \\
&=& 
\frac{2 \exp\left(-\frac{E}{\hbar\omega} - i\frac{Et}{\hbar} 
  + i\frac{\lambda x^2}{2\hbar} + i\frac{E}{\hbar\lambda}\ln \frac{|x|}{x_0}\right) }
  {(2\pi\hbar\omega E)^{1/4} (2\pi\hbar\lambda |x|)^{1/2}}
\nonumber \\
&&\times
  \frac{\sin \frac{\Delta E}{2\hbar} \left( t -\frac{1}{\lambda} \ln\frac{|x|}{x_0} -\frac{i}{\omega}\right)}
  {\frac{1}{2\hbar} \left( t -\frac{1}{\lambda} \ln\frac{|x|}{x_0} -\frac{i}{\omega}\right)}
\end{eqnarray}
At time $t$, the phases remain coherent within an energy band of width $\Delta E=\hbar/t$. The ETH approach rests on the fact that this band shrinks with time $t$. The coherence is controlled by the right-most sine function which for large times imposes the constraint $x\sim x_0e^{\lambda t}$ and by the term $\exp(i\lambda x^2/2\hbar)$ in the exponent.  The phase coherent bands in the variable $x^2$ are determined by the exponential, yielding the  position uncertainty
\begin{equation}
\Delta x \sim \frac{\hbar}{2\lambda x}  ~~.
\end{equation}
For the occupation probability of each coherent energy band this condition implies for large enough times $t$:
\begin{equation} 
\rho_E = \int_{-\infty}^{\infty} dx \Big| \Psi_{E,\Delta E} (x,t) \Big|^2
~=~
2\Delta E \frac{e^{-2E/\hbar \omega}}{\sqrt{2\pi\hbar \omega E}} ~~,
\end{equation}
which implies the normalization $\int_0^{\infty} dE ~\rho_E = \Delta E$.  Note the appearance of the thermal weight factor $e^{-2E/\hbar \omega}$, as predicted by the ETH.  This result can be easily extended to include the block structure in $x$:
\begin{equation} 
\rho(E,x)  = \frac{\Theta(\Delta x -|x-x_0e^{\lambda t}|)}{\Delta x}  ~ \frac{e^{-2E/\hbar \omega}}{\sqrt{2\pi\hbar \omega E}}
\end{equation}
with the normalization
\begin{equation}
\int_0^{\infty} dE~ \int_0^{\infty}dx~  \rho(E,x) = 1 ~~.
\end{equation}  
From the block matrix of all coherent energy--position bands $\rho(E,x)$ it is then straightforward to calculate the coarse grained entropy for large times:
\begin{eqnarray}
S &=&  \int_0^{\infty} dE~ \int_0^{\infty}dx~  [\rho(E,x) \ln \rho(E,x)] 
\nonumber \\
&\approx& \int_0^{\infty} dE~ \int_0^{\infty}dx~  [\rho(E,x)\ln ( e^{\lambda t})] ~=~  \lambda t  ~~,
\end{eqnarray}
thus leading to a constant entropy growth rate $dS/dt = \lambda$ in agreement with the results (\ref{eq:dSdt}) obtained in the Husimi formalism.

Apparently, dephasing of $(\Delta E,\Delta x)$ blocks has a similar effect as phase space smearing, at least in this toy example.  This suggests that the phase of linear entropy growth may be a generic feature of entropy production in quantum field theories.

\subsection{Entropy growth rate of non-Abelian gauge fields}
\label{subsect:SU2}

\begin{figure}[h]
\centerline{
\includegraphics[width=0.75\linewidth]{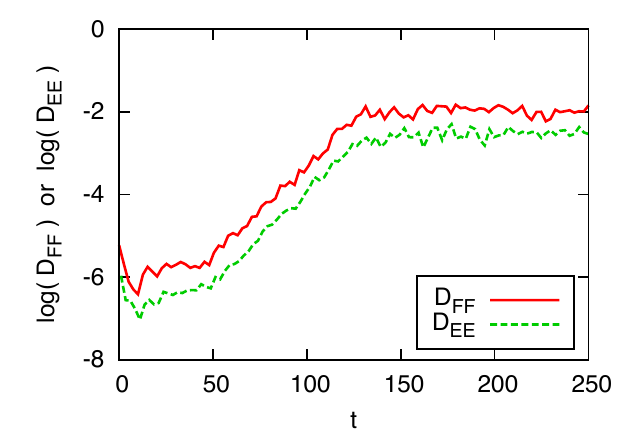}
}
\caption{
Time evolution of the distance in SU(2) simulation on a $4^3$ lattice for the two different distance measures.
All scales are given in the lattice unit.
%[From Kunihiro {\em et al.}\cite{Kunihiro:2010tg}]
}
\label{Fig:AY1}
\end{figure}

The analysis of the previous sections suggests that entropy production is, in general, dominated by a phase of linear growth with time, where the slope is given by the Kolmogorov-Sina\"i entropy growth rate (in short, the KS entropy). We assume that this is also true for highly excited quark-gluon systems like those created in high-energy heavy-ion collisions. In this case, the KS entropy can be calculated from the dynamics of {\em classical} gauge fields. Kunihiro {\em et al.}\cite{Kunihiro:2010tg} studied the real-time dynamics of classical Yang-Mills fields numerically after replacing continuous space by a three-dimensional cubic lattice. On this lattice they investigated the Hamiltonian 
\begin{eqnarray}
H&=&\frac{1}{2}\sum_{x,a,i}E_i^a(x)^2 + \frac{1}{4}\sum_{x,a,i,j}F_{ij}^a(x)^2 \ ,
\ ,\\
F_{ij}^a(x) &=& \partial_i A_j^a (x)-\partial_j A_i^a (x)
+ \sum_{b,c} f^{abc}A^b_i(x)A^c_j(x) \ ,
\end{eqnarray}
where $\partial_i$ denotes the central difference operator in the $i$-direction. The classical equations of motion are then
\begin{eqnarray}
\dot{A}_i^a(x) &=& E_i^a(x) \ ,\\
\dot{E}_i^a(x) &=& \sum_{j} \partial_j F_{ji}^a(x)+ \sum_{b,c,j} f^{abc}A^b_j(x) F_{ji}^c(x) \ .
\end{eqnarray}
The spectrum of Lyapunov exponents and the Kolmogorov-Sina\"i entropy were determined with two different distance measures.  As expected, the slope was found to be independent of the distance measure used, as shown in Fig.\ref{Fig:AY1}.

\begin{figure}[h]
\centerline{
\includegraphics[width=0.75\linewidth]{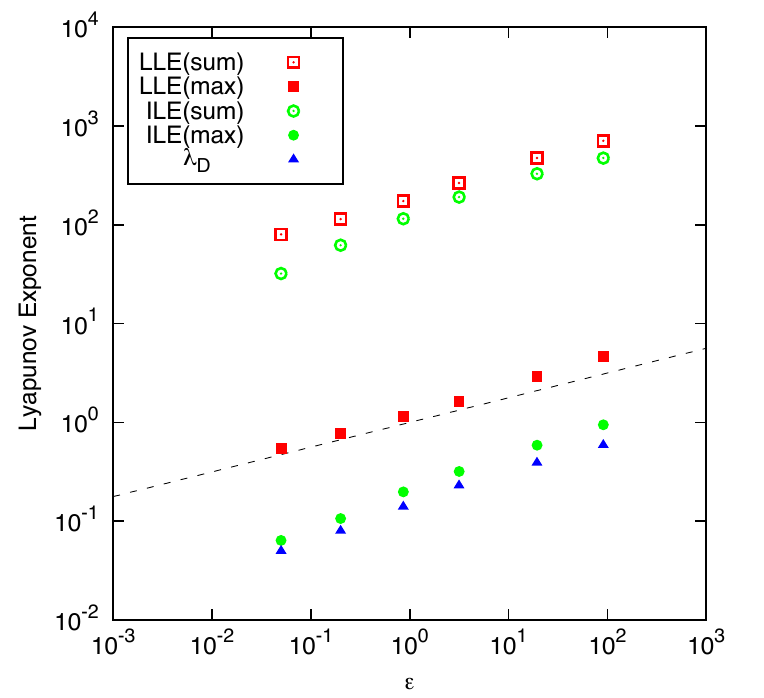}
}
\caption{The SU(3) results for various maximal and summed Lyapunov exponents.
The Kolmogorov-Sina\"i entropy corresponds to ILE(sum).  
The broken line is $\epsilon^{1/4}$.
}\label{Fig:AY6}
\end{figure}

In order to convert the results of the lattice simulations to dimensional, physical values, one needs to take into account that classical gauge theories have unphysical ultraviolet divergencies. This limitation of classical field theory is already encountered for point charges in classical electrodynamics.  The physical reason is that classical point charges do not exist; no particle can be localized better than within its de Broglie wave length.  Thus the lattice spacing $a$ is not an unphysical parameter which should be sent to zero as in lattice simulations of the euclidean quantum field theory.  Rather it determines the length scale below which neglected quantum effects become relevant. There are various ways to determine $a$.  For SU($N_c$) gauge theories the inverse damping rate for infrared modes is e.g.\cite{Braaten:1990it}
\begin{equation}
\frac{1}{\gamma} ~=~ \frac {24 \pi}{6.64g^2N_cT} ~\approx~ \frac{1}{T}
\end{equation} 
for $N_c=3$. Alternatively one can compare the energy density on the lattice and in the continuum 
\begin{eqnarray}
\epsilon_{\rm Lattice}(T) &=& 2(N^2-1)~\frac{T}{a^3}
\nonumber \\ 
\epsilon_{\rm Cont}(T) &=& 2(N^2-1) ~ \frac{\pi^2}{30} T^4
\nonumber \\
\Rightarrow ~~a&=& \Bigl( \frac{30}{\pi^2T^3} \Bigr)^{1/3} ~\sim~ \frac{1.4}{T}  
\end{eqnarray}
The fact that $a$ should scale like $\epsilon^{-1/4}$ implies that the entropy slope in lattice units should scale like $\epsilon^{1/4}$ which it does, as shown in Fig.\ref{Fig:AY6}.

\begin{figure}[h]
\centerline{
\includegraphics[width=0.75\linewidth]{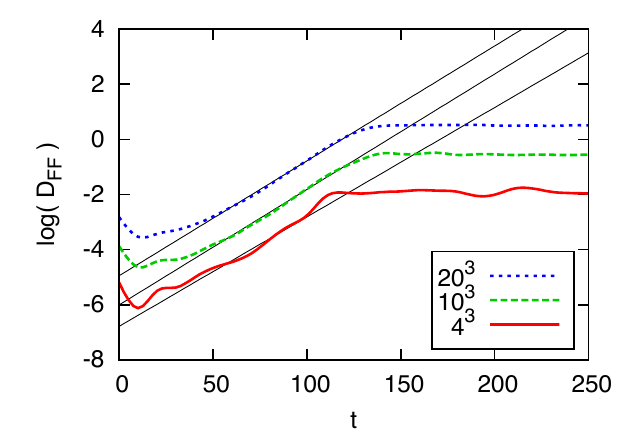}
}
\caption{
Time evolution in SU(2) simulation on $4^3, 10^3$, and $20^3$ lattices with the same energy density.
}\label{Fig:AY2}
\end{figure}

Kunihiro {\em et al.} also confirmed that the results for the Lyapunov exponents are volume independent, see Fig.\ref{Fig:AY2}.
Finally, in view of the discussion below it is worth mentioning that within classical Yang-Mills theory thermalization follows the bottom-up rather than the top-down scenario, i.~e.\ soft modes grow stronger and thermalize faster than hard modes.
More details can be found in Kunihiro {\em et al.}\cite{Kunihiro:2010tg}  The conclusion from these investigations is that the thermalization time in a relativistic heavy-ion collision must be approximately 2 fm/$c$ and can hardly be shorter than 1 fm/$c$. 

The success of the classical lattice simulation for the KS entropy of the non-abelian gauge field poses the question how generic this behaviour is.  The lattice regularized Yang-Mills equations describing the dynamics of classical color fields are known to be strongly chaotic,\cite{Muller:1992iw,Biro:1993qc} and the KS-entropy of the classical Yang-Mills field was shown to be a thermodynamically extensive quantity.\cite{Bolte:1999th}\  It could be that systems with less pronounced ergodic properties show a less clear-cut behaviour.  Even for quantum chromodynamics there are still many open issues.  For example, one can extend the lattice approach to include Gaussian fluctuations.\cite{Gong:1993fz}\   It would also be interesting to evaluate the increase of the coarse grained entropy associated with the growth of the initial quantum fluctuations around classical gauge fields in the colliding nuclei.\cite{Fukushima:2006ax}

Finally, let us comment on related work by Berges {\em et al.}\cite{Berges:2007re}  These authors analyzed the isotropization of the energy momentum tensor of classical Yang-Mills fields. In their Fig.~8,  low momentum modes show a behaviour reminiscent of that observed by Kunihoro {\em et al.} with a similar isotropization time of $1-2$ fm/$c$.   On the other hand, the unexpected behaviour of high momentum modes may not be physically significant, because these modes are subject to lattice artefacts.

\section{Entanglement Entropy}

\subsection{The basic concept}
\label{subsect:BC}

The {\em entanglement entropy} is defined as follows: Consider a quantum system $X$ composed of two complementary subsystems $B$ and $B'$ such  that $X = B \cup B'$, which is in a pure state $|X\rangle$. If the state of the subsystem $B$ is not a pure state, one says that the quantum states of $B$ and $B'$ are entangled. The density matrix of the subsystem $B$ can be written as $\rho_B = {\rm Tr}_{B'}\left(|X\rangle\langle X|\right)$, and the entanglement entropy is defined as
\begin{equation}
S(B) = - {\rm Tr}_B [\rho_B \ln\rho_B] .
\end{equation}
This definition measures the amount of entanglement of the quantum state of $B$ with the quantum state of $B'$. Obviously, for this definition to make sense $S(B)$ must be equal to $S(B')$ obtained by tracing the density matrix over $B$. The relationship $S(B)=S(B')$ implies that the entropy can only depend on the area of the common surface separating $B$ from $B'$, which we call $A(B) = A(B')$.  If the entanglement entropy is proportional to the surface area, as explicit calculations in tractable quantum field theories show, this implies that $S(B)=\kappa\, A(B) = \kappa\, A(B') =S(B')$, where $\kappa$ is the proportionality constant. Entanglement entropy thus contains by construction the holographic principle, making it one of the primary objects of interest for the thermodynamic properties of event horizons. The entanglement entropy was first quantitatively studied for a three-dimensional quantum field theory by Srednicki, who found that the value of the entropy depends quadratically on the ultraviolet momentum cut-off of the field theory.\cite{Srednicki:1993im}\ The concept trivially generalizes to spatial dimensions other than $d=3$.  

The one-dimensional case has been extensively studied for conformal field theories.\cite{Calabrese:2009qy}\  In that case, the entanglement entropy of an interval of length $\ell$ in the vacuum state of the field theory is given, up to a possible constant, by the general formula\cite{Holzhey:1994we,Vidal:2003xx}
\begin{equation}
S(\ell) = \frac{c}{3} \ln(\ell/a)  ~~,
\label{eq:S-entangled}
\end{equation}
where $c$ denotes the central charge of the conformal field theory and $a$ is the short-distance cut-off.  For this simple case, the analogue to the apparent conflict in three dimensions between the volume dependence of thermal entropy and surface dependence of entanglement entropy can be resolved as follows.  When the field is not in its vacuum state but in thermal equilibrium at $T>0$, then $S(B)$ measures not only the entanglement entropy with the surroundings of domain $B$, but also the thermal entropy contained in $B$.  As expected, the thermal contribution is proportional to the volume of $B$ (in one dimension simply the length $\ell$ of the interval).  The thermal generalization of (\ref{eq:S-entangled}) is\cite{Korepin:2003xx,Calabrese:2004eu}
\begin{equation}
S(\ell) = \frac{c}{3} \ln\left(\frac{1}{\pi a T} \sinh(\pi\ell T)\right) ,
\label{eq:S-thermal}
\end{equation}
which goes over into (\ref{eq:S-entangled}) in the limit $T\ell \ll 1$, and approaches the thermal entropy $S_{\rm th} = (c/3)\pi T \ell$ in the limit $T\ell \gg 1$. 

When the original vacuum state is perturbed by a quantum quench, the entanglement entropy increases with time and eventually reaches the thermal value. In the limit of a near critical one-dimensional conformal field theory on an interval of length $\ell$ one finds\cite{Calabrese:2009qy}
\begin{equation}
S(\ell,t) - S(\ell,0) \approx 
\frac{\pi c}{6\tau_0} \times \left\{ 
\begin{array}{lcl}
t & \qquad & t < \ell/2 \cr
\ell/2 & \qquad & t > \ell/2 ,
\end{array} \right.
\end{equation}
where $\tau_0$ is the so-called extrapolation length. The final entropy corresponds to the temperature $T=(4\tau_0)^{-1}$ Remarkably, the entanglement entropy reaches its equilibrium value after a finite time $\ell/2$, which implies that the entangled information travels away from the center of the interval at the speed of light.

The concept of entanglement entropy has an important application to black holes.  In the theoretical investigation of the properties of black holes it was realized early on that thermodynamic consistency requires black holes to have some form of entropy, which was suggested by Bekenstein\cite{Bekenstein:1973ur}\ to be proportional to its surface $S_{\rm BH}=A/4$.  The relation $dE=dM=TdS_{\rm BH}$, which holds when some object falls into the black hole, requires black holes to have a nonzero temperature, e.~g.\  $T=1/8\pi M$ for a Schwarzschild black hole, which is exactly the Hawking temperature.  That the latter is of quantum origin (notwithstanding the fact that some of its properties can be derived from thermodynamics)  underlines the fact that entropy cannot be formulated consistently in a classical theory, as usually discussed in connection with the Third Law of Thermodynamics.  Equating statistical entropy and thermodynamic entropy implies that the number of quantum states must be proportional to $A/4$. This relation implies that all information contained in a black hole is encoded in the area of its horizon, which is a special case of the holographic principle.\cite{tHooft:1993gx,Susskind:1994vu}\  The latter states that the number of degrees of freedom $N$, understood as the number of independent quantum states describing a region $B$ of space-time is bounded by the area $A(B)$ of its boundary.\cite{Simon:2011zz}\  Many detailed investigations suggest that this definition may, indeed, avoid all apparent paradoxes of information flow across an event horizon that have been constructed over the years. 

For our discussion, two aspects are worth highlighting:
\begin{itemize}
\item
The surface entropy of a horizon can be identified, up to a constant factor, with the {\em entanglement entropy}.
\item
There exists a fundamental difference between a static black hole and a dynamically created and evaporating one, as the entanglement entropy (which is often identified with the von Neumann entropy) can remain zero for the latter. Thus the unitarity of the boundary gauge field 
theory might find its counterpart in in the unitarity of the black hole formation and 
evaporation process, see e.~g.\ Takayanagi and Ugajin.\cite{Takayanagi:2010wp}
\end{itemize}

There is a growing consensus that the formation and evaporation of a black hole is a unitary process, at least in asymptotically flat or AdS space-times.\cite{Bousso:2002ju}  This implies that the von Neumann entropy stays constant throughout the process and 
information about the initial state can, in principle, be recovered from careful measurements of correlations within the Hawking radiation.\cite{Balasubramanian:2011dm}  The apparent (Bekenstein) entropy assigned to a black hole is an expression of the inability of a classical observer to distinguish between the many micro-states of the true quantum geometry.  In the ``fuzzball'' picture\cite{Mathur:2005zp}\ of the black hole, the horizon encloses the region in which the average micro-state deviates strongly from the classical geometry.  The finite size of the fuzzball is set by the uncertainty relation, which dictates that a micro-state carrying enough energy to represent the mass of the black hole will, in general, be spread out in space.  One can easily imagine that a phase space smearing of a micro-state, similar to the Husimi transform of the Wigner function, would map the the pure state of the quantum geometry into the entropy carrying, quasi-thermal geometry of a classical black hole.  The Bekenstein entropy 
would then have to be understood as the coarse grained entropy for a transitory black hole created by gravitational collapse as well as the entanglement entropy of an eternal black hole.

Takayanagi and Ugajin\cite{Takayanagi:2010wp}\ have argued that under certain circumstances the entanglement entropy of a transient black hole can remain zero. This could happen, e.g., if the forming and decaying black hole is microscopic so that it decays into a small number of excitations whose detailed phase relationship can be observed.  They construct a toy model in which ``black hole'' formation is a periodic process, which obviously does not entail entropy growth.  This is a special case of the general observation that the concept of entropy for a finite system only makes sense on time scales much shorter than the Poincar\'e recurrence time.  While this is practically true for most systems with many degrees of freedom, it does not apply to microscopic systems, and it may not apply to simplified systems that have been constructed to be mathematically tractable.

\begin{figure}[h]
\centerline{
\includegraphics[width=0.75\linewidth]{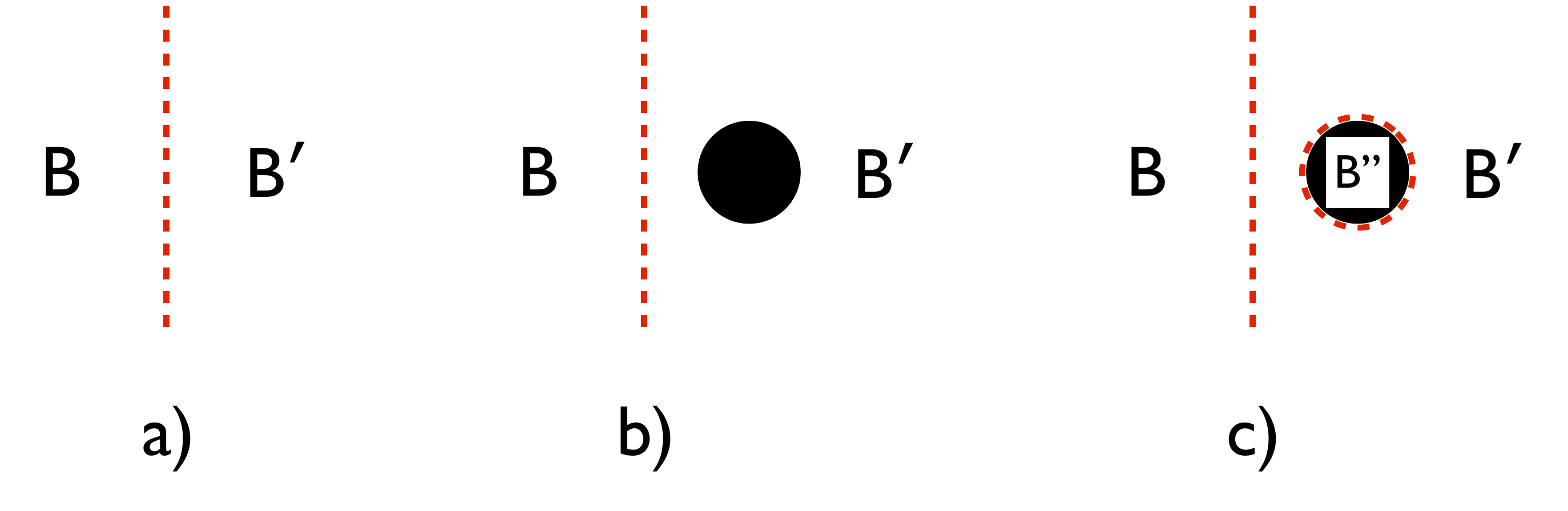}
}
\caption{An illustration of the meaning of unitarity within the formulation of entanglement entropy: a) two regions of space separated by a boundary; b) one region (B$'$) containing a transitory black hole; c) one region (B$'$) containing an eternal black hole, which can be considered as a third region of space (B$''$) with its event horizon as natural boundary with respect to region B$'$.}
\label{Fig:Entangle_1}
\end{figure}

We also note that the von Neumann entropy of an eternal black hole may be non-zero.  The fundamental difference between a stationary and a transitory black hole with respect to their entanglement entropy is illustrated in Fig.\ref{Fig:Entangle_1}.  Before the black hole is formed, $B$ and $B'$ cover the full quantum system  $X = B \cup B'$.  Thus entanglement entropy can be  defined as the minimal surface between $B$ and $B'$.  If a black  hole develops inside of $B'$ this minimal surface does not change at all (case $b$). As the process is unitary, $B'$ undergoes some non-trivial internal reordering of states, but this does not affect the surface between $B$ and $B'$.  However, in the presence of a stationary black hole,  $X \neq B \cup B'$ at early times and one has to define entanglement entropy from the minimal surfaces between the three regions $B$, $B'$ and $B''$ (case $c$).

\subsection{The AdS/CFT Duality}

The reason we discussed the question of entropy of a black hole at some length is that black holes play an important role in holographic gravity duals of interacting quantum field theories.  The {\em AdS/CFT duality}\cite{Maldacena:1997re,Aharony:1999ti}\  posits that superstring theory in AdS$_5\times S^5$ is holographically dual to a conformal field theory in Minkowski space, the ${\cal N}=4$ supersymmetric $SU(N_c)$ gauge theory.   In the limit of a large number of colors $N_c$ the string theory approaches the classical limit of the supergravity theory on AdS space.  When a black hole with Schwarzschild radius much larger than the AdS curvature radius (a ``black brane'') is added to the AdS$_5$ space, the string theory becomes dual to the thermal super-Yang-Mills theory in Minkowski space.  The connection to relativistic heavy-ion collisions originates in the observation that for very high temperatures, i.~e.\ deeply in the deconfined phase, QCD is approximately conformal and fermionic contributions to thermal properties should be much less important than gluonic ones for large $N_c$.  In addition, supersymmetry is broken by finite temperature, and the adjoint superpartners of the gauge field (``gauginos'') become effectively massive and decouple from the infrared sector of the theory.  Thus the absence of ${\cal N}$-fold supersymmetry in thermal QCD should be less relevant than in the vacuum theory.  Finally, the question whether the thermodynamic properties of $SU(N_c)$ gauge theories already have reached the large-$N_c$ limit for $N_c=3$ has recently been answered in the affirmative by lattice-QCD simulations.\cite{Panero:2009tv} 

For those readers who are not familiar with AdS/CFT duality, we now give a very brief description and refer to Natsuume\cite{Natsuume:2007zz}\ for more details and references.  As already stated, AdS/CFT duality establishes a correspondence between superstring theory in the AdS$_5\times S^5$ space-time and the ${\cal N}=4$ supersymmetric $SU(N_c)$ gauge theory in Minkowski space. For many purposes, the $S^5$ part of the ten-dimensional space-time can be ignored.  The correspondence is thus between a five-dimensional theory of quantum gravity and a four-dimensional gauge theory.  The most important aspect of the correspondence is that the strong 't Hooft coupling limit ($g^2N_c \to \infty$) of the gauge theory is mapped onto the weak coupling limit of the string theory, which is well described by its classical limit, the supergravity theory on space-times with asymptotic AdS$_5$ geometry.  The field theory can be considered as ``living'' on the asymptotic boundary of the AdS$_5$ space-time; the AdS$_5$ space-time is therefore usually referred to as the ``bulk''.  

There is an extensive dictionary of correspondences between expectation values of operators in the field theory and geometrically defined objects in the dual supergravity theory.\cite{Aharony:1999ti}\  For example, the expectation values of some local operators of the gauge theory, such as the energy-momentum tensor, are related to the metric or other local fields in the gravity theory.  The expectation values of non-local operators in the gauge theory, such as two-point functions or Wilson loops, correspond to the actions associated with geodesics, minimal surfaces, or other geometric objects in the dual supergravity theory.\cite{Maldacena:1998im} 

The thermal state of the field theory, i.~e.\ the equilibrated gauge field plasma, is represented by Anti-deSitter space with an imbedded black hole of mass $M$, whose Hawking temperature $T_H = M^{1/4}/\pi$ equals the physical temperature $T$ of the field theory.  Empty Anti-deSitter space corresponds to the vacuum state of the field theory.  Thermalization of the theory is thus dual to the process of black hole formation in the bulk.  The energy deposition occurring in a heavy ion collision corresponds to the injection of energy in the outer regions of the bulk, which then falls by gravitational attraction and eventually forms a black hole, providing a holographic description of the process of thermalization in the gauge theory.

In recent years it has been understood that AdS/CFT duality makes it possible to obtain rigorous predictions for strongly coupled gauge theories at high temperature which seem to agree with phenomenological results deduced from high-energy heavy-ion collision data.\cite{CasalderreySolana:2011us}\  However, it is important to keep in mind that QCD is not a strongly coupled theory in the high temperature limit, where it becomes approximately conformal.   On the other hand, in the temperature range of practical interest ($T \leq 3 T_c \approx$ 500 MeV) where thermal QCD is a strongly coupled theory, it is far from being conformally invariant. Lattice-QCD simulations show that the trace anomaly of the energy-momentum tensor, $T^\mu_\mu = \varepsilon - 3P$, is large and of the same order as $\varepsilon$ itself (see Borsanyi {\em et al.}\cite{Borsanyi:2010cj}, Fig.~6).  It is thus not obvious that the AdS/CFT correspondence in its simplest form is a good model for thermal QCD in the region of interest.  This has motivated attempts to find improved holographic duals  (see, e.~g., Alanen {\em et al.}\cite{Alanen:2010tg}, Kajantie {\em et al.}\cite{Kajantie:2011nx}).

\subsection{Holographic Entanglement Entropy}

\begin{figure}[h]
\centerline{
\includegraphics[width=0.5\linewidth]{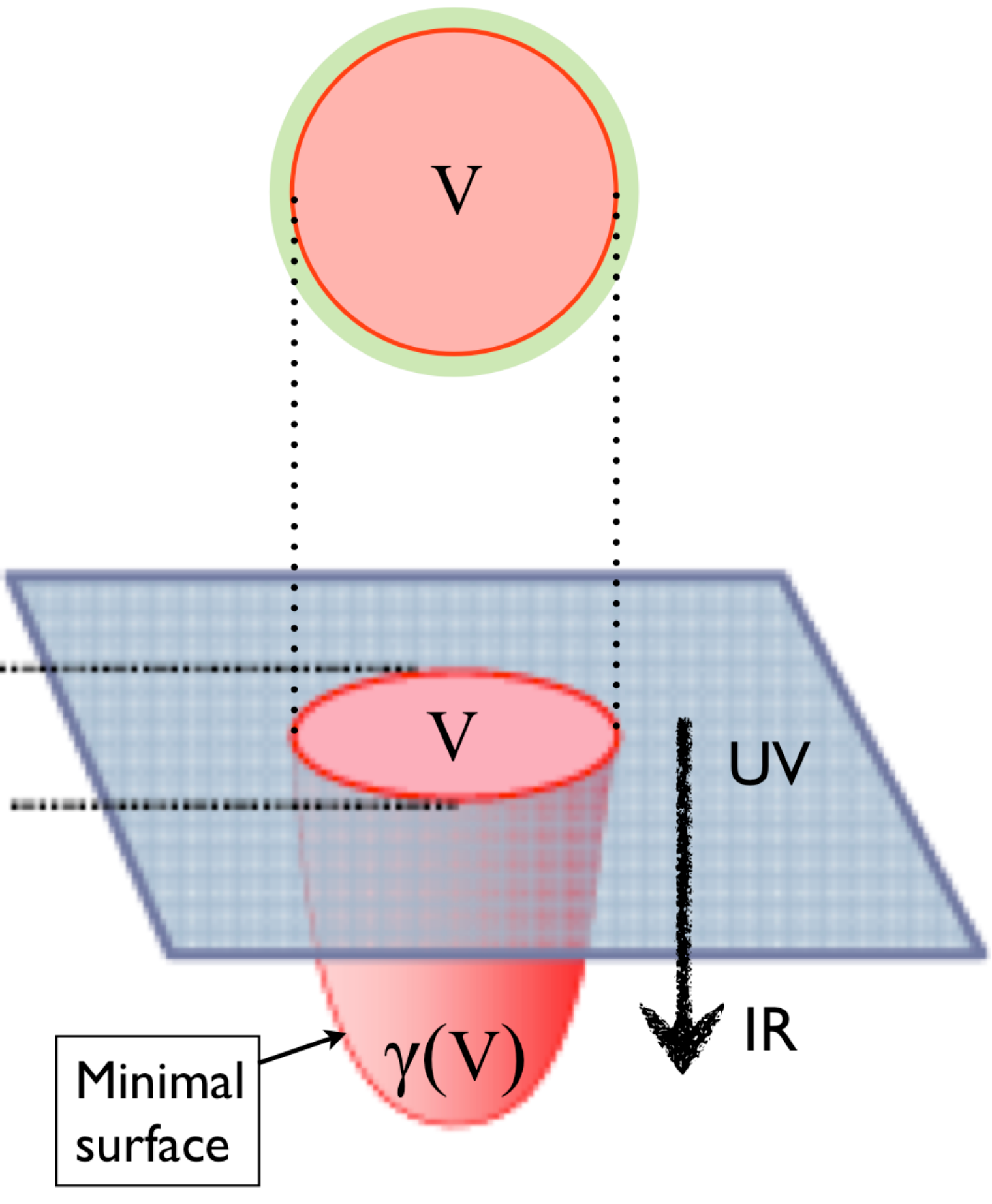}
}
\caption{Holographic representation of the entanglement entropy of a finite region $V$ of space as the area of the minimal hypersurface $\gamma(V)$ in the bulk, which has the same boundary as $V$.}
\label{Fig:EE}
\end{figure}

A volume of Minkowski space $V$ filled with a quantum field in its vacuum state has a non-vanishing entanglement entropy associated with it, because the field modes of its complement necessarily leak across the boundary (see Fig.~\ref{Fig:EE}).  Although the field is known to be in the vacuum state within $V$, nothing is assumed to be known about the state of the field outside.  As mentioned in Section~\ref{subsect:BC}, this ultraviolet divergent entanglement entropy is proportional to the surface area of the volume.\cite{Srednicki:1993im}\  A holographic dual description of the entanglement entropy of a region of Minkowski space-time was proposed by Ryu and Takayanagi\cite{Ryu:2006bv,Ryu:2006ef}\ (see Nishioka et al.\cite{Nishioka:2009un}\ for a review).  They extended the surface of the volume into dimension of the bulk to form a minimal hypersurface $\gamma(V)$, whose boundary coincides with the surface of the Minkowski space volume $V$.  This is illustrated in Fig.~\ref{Fig:EE}.  The entanglement entropy associated with this surface is defined as:\cite{Ryu:2006bv,Ryu:2006ef}
\begin{equation}
S_V = \frac{||\gamma(V)||}{4 G_N}
\label{eq:RT}
\end{equation}
where $||\gamma(V)||$ denotes the volume of $\gamma(V)$, which has the same dimensionality as $V$, and $G_N$ is Newton's gravitational constant.  When the field is in the vacuum state, $\gamma(V)$ extends into pure AdS$_5$ space, and one finds that its ultraviolet divergent volume is proportional to the surface area of $V$. When the Anti-deSitter space contains a black hole and the region $V$ is large enough, the hypersurface $\gamma(V)$ hugs the event horizon of the black hole, and $||\gamma(V)||$ acquires an additional contribution proportional to the covered horizon area, which is proportinal to the volume of $V$. By explicit calculation one finds that the additional contribution to the entanglement entropy is precisely the thermal entropy of the gauge field in the Minkowski space region $V$.  Ryu an Takayanagi confirmed the equivalence between the entanglement entropy calculated in the field theory and that calculated holographically by means of (\ref{eq:RT}) in several other tractable cases, but no explicit equivalence proof is known.  In the following, we assume that the equivalence holds, in general. 
 
The relation (\ref{eq:RT}) is remarkable in that it relates the Bekenstein entropy of black holes to the usual thermal entropy of a field theory by holographic duality.  We can imagine applying this concept to a relativistic heavy ion collision, in which a certain region of space is suddenly filled with high energy that quickly equilibrates into a hot quark-gluon plasma, which then expands, cools, and eventually dissolves. In the dual description, energy is injected asymptotically into an AdS$_5$ space, collapses to form a black hole (or black brane), which eventually evaporates by Hawking radiation.  The formation of the black hole, i.~e.\ the dual of the thermalization process, can be tracked by the growth of the entanglement entropy defined in (\ref{eq:RT}).  However, in this context, the holographic duality of the entanglement entropy concept raises a number of questions:
\begin{itemize}
\setlength{\itemsep}{0pt}
\item
How is the entropy of a black brane in AdS$_5$ related to the von Neumann and Wehrl entropies in the boundary gauge theory?
\item
Can the duality be extended to the dynamical process of black brane formation in AdS$_5$ and thermalization in the boundary field theory?  
\item 
Is it correct to apply the entanglement entropy correspondence, which has been verified to be dual in the case of thermal equilibrium, to non-equilibrium situations?
\end{itemize}
None of these questions can be conclusively answered at present. Nevertheless, we want to review some of the ideas and insights which might contribute to find relevant answers in the future. Also we will review results from one specific model study. 

A crucial element in the discussion of thermalization is the form the fluctuation-dissipation theorem takes in the bulk.  Following Caron-Huot {\em et al.}\cite{CaronHuot:2011dr}\ one can argue as follows:  The large black brane in AdS emits Hawking radiation in the same way as black holes in Minkowski space. This does not lead to a heating of the boundary, because perturbations along outgoing geodesics are reflected at the boundary and, therefore, end up again in the black brane. However, the quantum fluctuations associated with this Hawking radiation induce fluctuations in the boundary, which by the fluctation-dissipation theorem are related to dissipation, and push the system towards equilibrium.  The closer a fluctuation is located to the horizon, the longer it takes to reach the boundary and the stronger it will be red-shifted, while all higher momentum fluctuations get reabsorbed by the black brane on a time scale of order $1/T$.  In the boundary field theory, therefore, ultraviolet modes thermalize first, infrared modes last.  

To gain more insight into this complex situation one studies the reaction of AdS/CFT for various types  of sudden violent perturbations, often called {\em quantum quenches}.  The entanglement entropy then traces the evolution from such a highly excited, but highly phase correlated initial state to the state of thermal equilibrium.\cite{Calabrese:2009qy}\  To reduce the technical difficulty of such calculations, one often studies lower dimensional theories.  In the next section we will review results in 2, 3, and 4 dimensions for a schematic model of black hole formation in the bulk.\cite{Albash:2010mv,Balasubramanian:2010ce,Balasubramanian:2011ur}

\subsection{Thermalization}

Albash {\em et al.}\cite{Albash:2010mv} and Balasubramanian {\em et al.}\cite{Balasubramanian:2010ce,Balasubramanian:2011ur}\ considered the case where a finite energy density was deposited in the bulk near the boundary at a certain initial time, resulting in a shell of null dust falling in the fifth dimension until it forms an event horizon and a black brane is created. The resulting metric was then probed by strings and membranes with endpoints on the boundary. Thermalization was studied as a function of the Minkowski space separation between the endpoints, respectively the area of the Wilson loop, as a function of time.  The time required for the geodesic length or the membrane area to reach that of the black brane equilibrium configuration was interpreted as an estimate for the physical thermalization time, as measured by the entanglement entropy.  The results of this study showed that complete thermalization occurs within a time that is equal to half of the spatial diameter of the probe in the boundary space. 
 
\begin{figure}[h]
\centerline{
\includegraphics[width=0.75\linewidth]{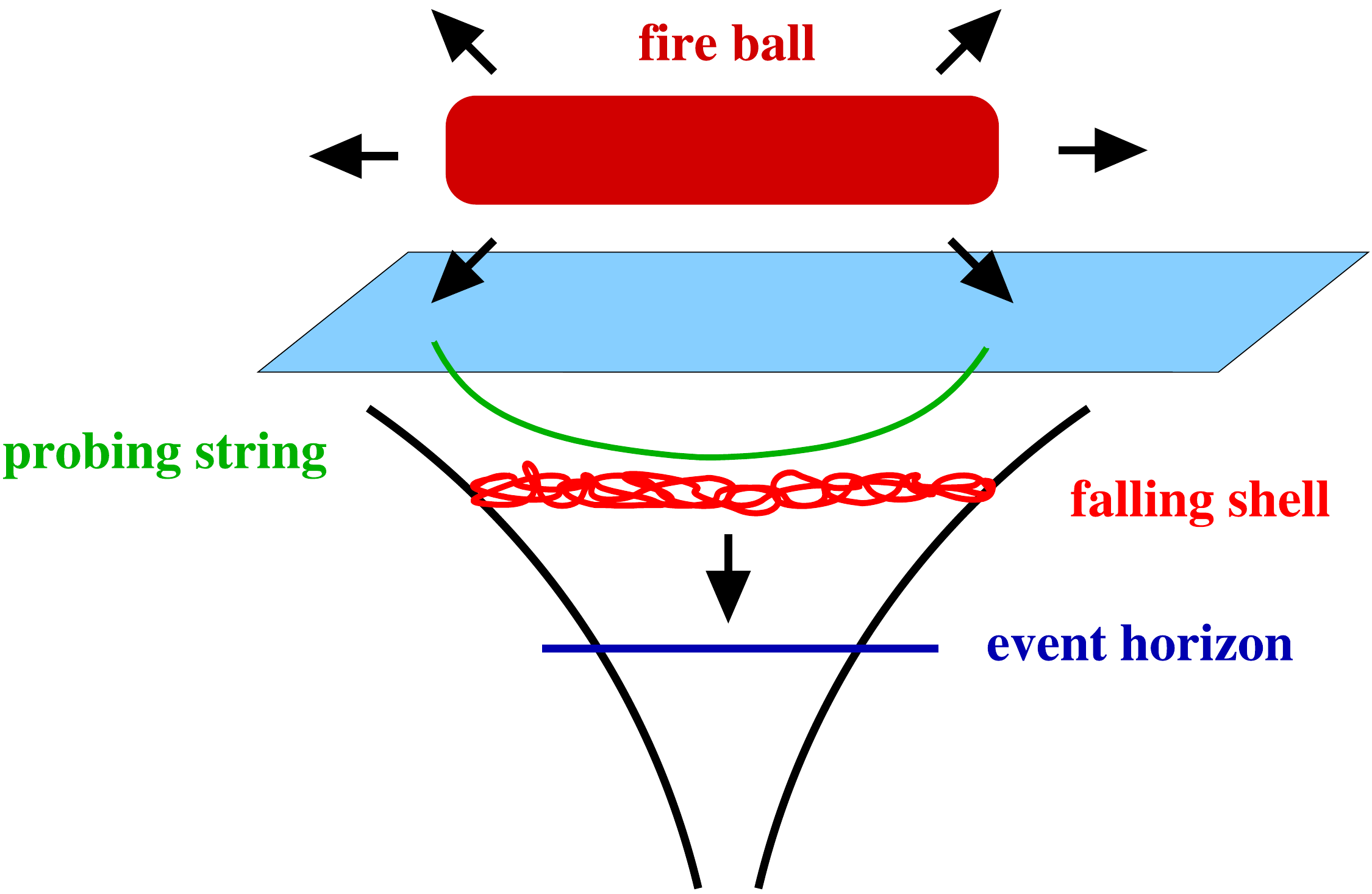}
}
\caption{Illustration of the case studied. Energy is deposited on the boundary at a certain starting time. 
This situation should be seen as model for the creation of a fireball in a high-energy heavy ion collision.
This energy deposition leads to the propagation of a shell of what is called 'null dust' in the fifth dimension.
Its dynamics can be solved analytically and absorbed into a specific metric, the Vaidya metric. The latter is then 
probed with e.g. a string with fixed endpoints on the boundary, which corresponds to the correlation function between 
two operators of very large mass-dimension. The analytic form of the geodetic is known for the equilibrium case after 
the shell moved beyond the horizon. The time scale with which the geodetic length approaches its value 
for that solution is identified with the thermalization time.}
\label{Fig:string_1}
\end{figure}

The AdS space dual to the thermalizing field theory is probed by geodesics and surfaces with endpoints on the boundary, see Fig.\ref{Fig:string_1}.  The effect of the in-falling shell can be expressed geometrically in so-called Poincar\'e coordinates by the Vaidya metric
\begin{equation}
ds^2 = \frac{1}{z^2}\left[-\left(1 - m(v)z^d\right) dv^2 - 2 dz\, dv + d\mathbf{x}^2 \right] \,,
\label{eq:Vaidya}
\end{equation}
where $v$ labels in-going null trajectories and the AdS radius has been set equal to unity. The boundary space-time is located at $z=0$ and $\mathbf{x}=(x_1,\dots,x_{d-1})$ correspond to the spatial coordinates on the boundary.  The mass function of the in-falling shell was taken to be:
\begin{equation}
m(v)=\frac{M}{2}\,\left( 1+\tanh\frac{v}{v_0} \right) ,
\label{eq:mv}
\end{equation}
where $v_0$ parametrizes the thickness of the shell falling along $v=0$, which is assumed to be very small.\cite{Lin:2006rf,Lin:2007fa,Lin:2008rw,Bhattacharyya:2009uu}\  For a sufficiently large separation of the two endpoints on the boundary and before the shell has formed an event horizon, the geodesic curve in the bulk connecting the endpoints on the AdS boundary will ``punch'' through the shell and explore the still un-thermalized AdS geometry below the shell.  The larger the portion of the curve extending below the shell, the larger is the difference between the geodesic length of the curve and that of a geodesic connecting the two endpoints in the AdS-Schwarzschild geometry, which represents full thermalization. 

\begin{figure}[h]
\begin{center}
\includegraphics[width=0.6\textwidth]{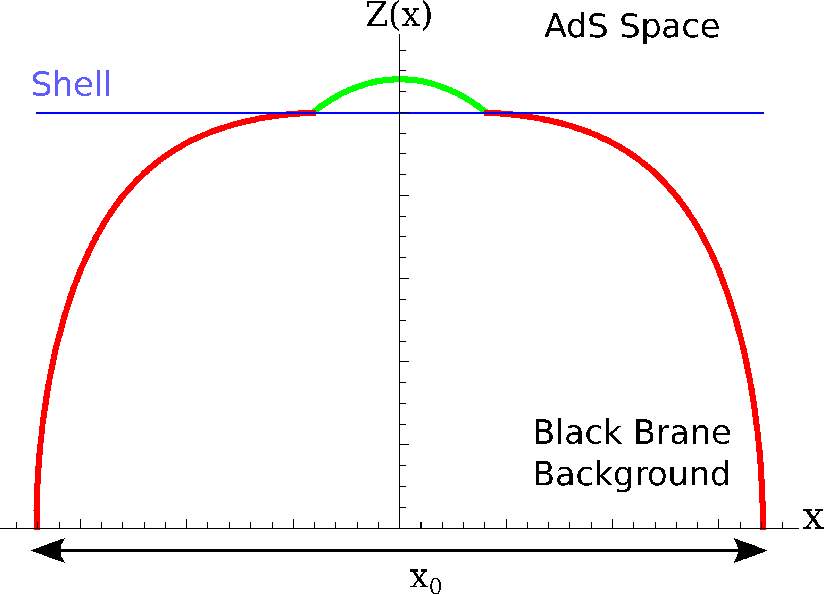} 
\caption{Example of a space-like geodesic that starts and ends on the 
boundary of AdS ($z=0$) with a separation $x_0$.  Outside the shell, the geodesic 
propagates in a black brane geometry, while inside it propagates in an empty AdS geometry.  
	%%(from  \cite{Balasubramanian:2011ur}).
}
\label{FigCrossingTheShell}
\end{center}
\end{figure}

It is then possible to obtain analytic or numerical solutions for different dimensions and probes.   Some results were obtained by Balasubramanian {\em et al.}\cite{Balasubramanian:2010ce,Balasubramanian:2011ur}\ for the entanglement entropy.  In one spatial dimension ($d=2$) the entanglement entropy is holographically given by the length of the bulk geodesic connecting the two endpoints of the linear region in the boundary space. In two spatial dimensions ($d=3$) the entanglement entropy is holographically given by the minimal area of the bulk surface, whose perimeter coincides with the perimeter of the circular boundary area.  In three spatial dimensions ($d=4$) the entanglement entropy is holographically given by the minimal volume of the bulk hypersurface, whose surface coincides with the surface of the spherical boundary region.  

\begin{figure}[ht]
\centering
\includegraphics[height=4cm]{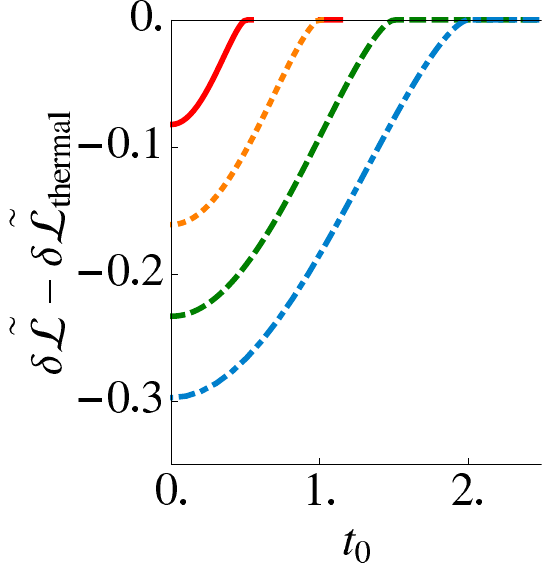} \hfil
\includegraphics[height=4cm]{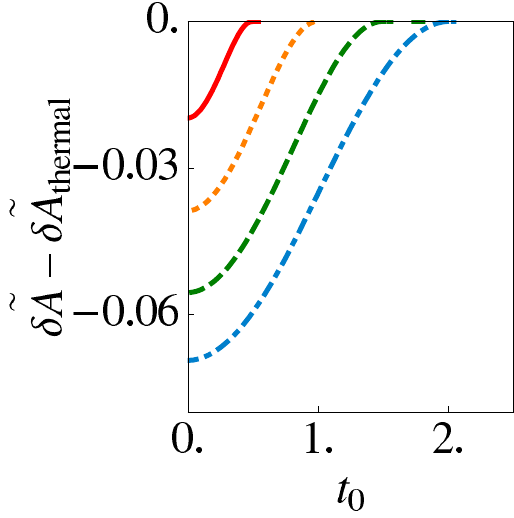} \hfil
\includegraphics[height=4cm]{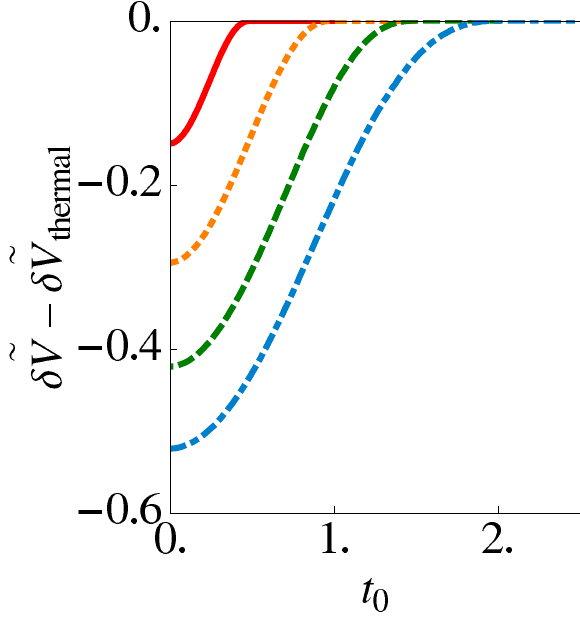}
\caption{The difference in normalized geodesic length $\Delta(\delta{\tilde {\cal L}})$, minimal surface area $\Delta(\delta\tilde{{\cal A}})$, and minimal volume $\Delta(\delta\tilde{{\cal V}})$, respectively, to that of the thermalized state as a function of  boundary time $t_0$ for dimensions $d=2$ (left), $d=3$ (center), and $d=4$ (right).  The results are for a thin in-falling shell ($v_0 = 0.01$).   The boundary separations were taken to be $\ell = 1,2,3,4$ and radii $R = 0.5, 1, 1.5, 2$ (from top to bottom curve), respectively.  All quantities are given in units of $M$.  Complete thermalization occurs precisely at $t_0=\ell/2$ or $t_0=R$, respectively.
	%%(from  \cite{Balasubramanian:2011ur}).
}
\label{fig:deltaSE}
\end{figure}

The results are shown in Fig.~\ref{fig:deltaSE}.  In each dimension, the difference between the minimal length (or area or volume) in the bulk for the geometry at a given time and minimal length (area, volume) in the fully thermalized AdS-Schwarzschild geometry is shown, normalized to the boundary volume ($\tilde{\cal L} = {\cal L}/\ell$, $\tilde{\cal A} = {\cal A}/(\pi R^2)$, $\tilde{\cal V} = {\cal V}/(4\pi R^3/3)$).  In every case one finds that thermalization as measured by the entanglement entropy is achieved within the time given by precisely half the diameter of the probe in the boundary divided by the speed of light.  The main results of this investigation are:\cite{Balasubramanian:2010ce,Balasubramanian:2011ur}
\begin{itemize}
\setlength{\itemsep}{0pt}
\item Thermalization is an extremely fast process in the strongly coupled gauge theory, which is only constrained by causality. The information loss from a finite region proceeds at the speed of light. This is especially remarkable because the strongly coupled gauge theory does not support quasi-particle excitations, with the exception of phonons, which do not play a role in this process.  The mechanism by which the information is transported out of the thermalizing volume is unknown.  However, the result suggests that the strongly coupled gauge theory behaves like a {\em fast scrambler} as defined by Sekino and Susskind.\cite{Sekino:2008he}
\item
Other observables reach their equilibrium values faster than the entanglement entropy. This is not surprising, because the entropy is sensitive to all degrees of freedom of the gauge theory, while other observables are sensitive only to a subset.
\item
Short distances thermalize first, i.e. thermalization occurs top-down for a gauge theory in the limit of very strong coupling, while it occurs bottom-up for weakly coupled gauge theories.\cite{Baier:2000sb}
\end{itemize}

\section{Summary}

The problem of thermalization of highly excited states in quantum chromodynamics and gravity has been intensively studied during the past decade. In QCD, the interest was driven by the experimental results from relativistic heavy ion collisions, which indicated that a thermal quark-gluon plasma is formed on an extremely short time scale of order 1 fm/$c$. In gravity theories, the main motivation was the desire to resolve the information paradox associated with black hole formation. In both situations, the question can be phrased in terms of the growth rate of an appropriately defined entropy. While the von Neumann entropy of a closed system does not change with time in any quantum theory due to the unitarity of the time evolution, other measures of entropy, which take into account the principal or practical inability of an observer to measure all details of a system, can grow with time and eventually approach the thermal equilibrium value.

In this review, we covered several different definitions of such a coarse grained entropy with broad applicability: Husimi's phase-space smearing of the Wigner distribution, de-phasing of the quantum state underlying the eigenstate thermalization hypothesis, and entanglement entropy applied to holographic duals of the quantum system. We showed that de-phasing and Husimi smearing result in the same entropy growth rate in the case of a single unstable mode, which was equal to the KS entropy describing the growth rate of the coarse grained entropy in the classical limit. This observation motivated a numerical study of the entropy growth rate of the non-abelian SU(2) gauge theory, reviewed in Section~\ref{subsect:SU2}, which confirmed the rapid thermalization observed in the experiments. 

The concept of entanglement entropy and its holographic equivalent enables rigorous calculations of the approach to thermalization in the strongly coupled conformal supersymmetric gauge theory. In this approach, the process of thermalization is dual to the formation of a black hole, linking the study of thermalization in the $(3+1)$--dimensional gauge theory to the study of black hole formation in $(4+1)$--dimensional supergravity. The concept of entanglement entropy for a bounded region of space parallels that of the Husimi smeared entropy (the Wehrl entropy), because it accounts for the position uncertainty of field modes with a given momentum. The two concepts differ in that the entanglement entropy only accounts for the uncertainty relation smearing of field modes near the boundary, while the Wehrl entropy implements the uncertainty smearing of all field modes. The holographic duality also provides compelling evidence for the view that the formation and subsequent evaporation of a black hole must be a unitary process in a quantum theory of gravity, which does not result in an increase of the von Neumann entropy, but is accompanied by a maximal increase in an appropriately defined coarse grained entropy.

Notwithstanding these significant theoretical advances, there remain many unresolved questions. How exactly are the different notions of coarse grained entropy connected? When does black hole formation and decay result in an increase in the coarse grained entropy, and when does it represent a practically reversible process as in the toy model constructed by Takayanagi and Ugajin?\cite{Takayanagi:2010wp}\ How do the same concepts apply to the cosmic Big Bang? We hope that this review will motivate some of its readers to study and ultimately resolve such questions.

\section*{Acknowledgments} 

This work was supported in part by grants from the U.~S.\ Department of Energy (DE-FG-05ER-41367) and the German BMBF.


%\bibliographystyle{h-physrev5}
%bibliography{BigPic}

\end{document}